\documentclass[aps,prd,superscriptaddress,preprintnumbers,showkeys,twocolumn]{revtex4-1}
\usepackage{graphicx,float,wrapfig,subfigure}
\usepackage{amsfonts,amsmath,amssymb,amstext}
\usepackage{latexsym}
\usepackage{bm}
\usepackage{color}
\usepackage[normalem]{ulem}
\usepackage{MnSymbol}
\usepackage[colorlinks=true,linktocpage=true,linkcolor=blue,citecolor=blue]{hyperref}
\usepackage{slashed}

\definecolor{red}{rgb}{0.7,0,0}
\definecolor{green}{rgb}{0,0.5,0}

\begin{document}
 
 \title{Anisotropic charge transport in strongly magnetized relativistic matter}
 
 \author{Ritesh Ghosh}
 \email{Ritesh.Ghosh@asu.edu}
 \affiliation{College of Integrative Sciences and Arts, Arizona State University, Mesa, Arizona 85212, USA}

 \author{Igor A. Shovkovy}
 \email{igor.shovkovy@asu.edu}
 \affiliation{College of Integrative Sciences and Arts, Arizona State University, Mesa, Arizona 85212, USA}
 \affiliation{Department of Physics, Arizona State University, Tempe, Arizona 85287, USA}
 
 \date{November 19, 2024}
 
\begin{abstract}
We investigate electrical charge transport in hot magnetized plasma using first-principles quantum field theoretical methods. By employing Kubo's linear response theory, we express the electrical conductivity tensor in terms of the fermion damping rate in the Landau-level representation. Utilizing leading-order results for the damping rates from a recent study within a gauge theory, we derive the transverse and longitudinal conductivities for a strongly magnetized plasma. The analytical expressions reveal drastically different mechanisms that explain the high anisotropy of charge transport in a magnetized plasma. Specifically, the transverse conductivity is suppressed, while the longitudinal conductivity is enhanced by a strong magnetic field. As in the case of zero magnetic field, longitudinal conduction is determined by the probability of charge carriers to remain in their quantum states without damping. In contrast,  transverse conduction critically relies on quantum transitions between Landau levels, effectively lifting charge trapping in localized Landau orbits. We examine the temperature and magnetic field dependence of the transverse and longitudinal electrical conductivities over a wide range of parameters and investigate the effects of a nonzero chemical potential. Additionally, we extend our analysis to strongly coupled quark-gluon plasma and study the impact of the coupling constant on the anisotropy of electrical charge transport.
\end{abstract}


 \maketitle
 
 \section{Introduction}
 \label{Introduction}
 
 A relativistic plasma is a fundamental state of matter subject to very high temperatures or densities, where the characteristic energies of particles are comparable to or exceed their rest mass energies. Although achieving such extreme conditions in a laboratory is challenging \cite{Sarri:2015jyr,Chen_2023}, they are relatively common inside and around compact stars \cite{Sturrock:1971zc,Ruderman:1975ju,Arons:1983aa}. For example, the magnetospheres of magnetars are predominantly composed of electron-positron plasma \cite{Turolla:2015mwa,Kaspi:2017fwg}. Other forms of relativistic plasmas existed in the early Universe \cite{Grasso:2000wj}. One of them is the strongly interacting quark-gluon plasma (QGP), which was filling the early Universe until about 10 milliseconds after the Big Bang \cite{Yagi:2005yb}. In laboratory settings, small droplets of QGP are produced in relativistic heavy-ion collisions \cite{STAR:2005gfr,PHENIX:2004vcz,PHOBOS:2004zne}.
 
 Relativistic plasmas are often strongly magnetized. In magnetospheres of magnetars, plasma formation itself results from the strong magnetic field. The primordial plasma in the early Universe was also highly magnetized \cite{Grasso:2000wj,Durrer:2013pga,Vachaspati:2020blt}, and the corresponding seed fields provide a natural explanation for the presence of nonzero magnetic fields on intergalactic scales. In heavy-ion collisions, sizable magnetic field are generated by the electrical currents produced by colliding positively charge ions \cite{Skokov:2009qp,Voronyuk:2011jd,Deng:2012pc,Bloczynski:2012en,Guo:2019mgh}. 
 
 This study aims to investigate the electrical conductivity of magnetized relativistic plasma using first-principles quantum field theory. While  conductivity is a key transport characteristic, our understanding of it in the presence of strong background fields is incomplete. For plasmas in a weak magnetic field, it was studied in some detail in Refs.~\cite{vanErkelens:1984,Pike:2016aa}. For QGP in a quantizing magnetic field, several studies have been conducted using analytical methods \cite{Hattori:2016cnt,Hattori:2016lqx,Fukushima:2017lvb,Fukushima:2019ugr} and lattice calculations \cite{Buividovich:2010qe,Buividovich:2010tn,Astrakhantsev:2019zkr,Almirante:2024lqn}. Additionally, some estimates of electrical conductivity have been obtained from holographic models of quantum chromodynamics (QCD)  \cite{Mamo:2012kqw,Fukushima:2021got} and other phenomenological models \cite{Nam:2012sg,Kerbikov:2014ofa,Satapathy:2021cjp,Satapathy:2021wex,Bandyopadhyay:2023lvk}. Many authors have also used a kinetic theory approach to study the electrical conductivity of magnetized relativistic plasmas \cite{Kurian:2017yxj,Das:2019ppb,Thakur:2019bnf,Dey:2020awu}. However, due to the inherent limitations of kinetic theory, it does not provide deep insights into the regime of strong fields. Moreover, many studies rely on the relaxation-time approximation, which lacks detailed knowledge of the collision integral.
 
 Leveraging recent progress in developing quantum-field theoretic methods that utilize the Landau-level representation for Green's functions \cite{Wang:2021ebh,Ghosh:2024hbf}, in this study we investigate the anisotropic charge transport in strongly magnetized relativistic plasmas. We rely on Kubo's formalism to calculate the electrical conductivity within a gauge theory framework. We use a neutral quantum electrodynamics (QED) plasma, composed of equal numbers of electrons and positrons, as the primary testing ground for our study. We rely on the previously obtained fermion damping rate in the Landau-level representation to calculate the dependence of electrical conductivity on temperature and magnetic field. As we showed, at leading order in coupling, the corresponding damping rate is driven by quantum transitions between Landau-level states of charge carriers, which are accompanied by emission or absorption of photons. The conductivity results are also extrapolated to the case of strongly coupled QGP. We should mention that, with some minor modifications, the study can be extended to other types of relativistic plasmas. Some of our key results were briefly outlined in Ref.~\cite{Ghosh:2024fkg}.
 
 The paper is organized as follows. We start from the definition of the electrical conductivity tensor in terms of the current-current correlator in Sec.~\ref{sec:ElectricalConductivity}. By making use of a spectral representation for the fermion propagator, we relate the electrical conductivity to the damping rate of charge carriers. In Sec.~\ref{sec-Damping-rate}, we discuss the damping rate in the Landau-level representation that was derived in an earlier study \cite{Ghosh:2024hbf}. The damping rate, which was obtained at the leading order in coupling, takes into account the dependence on the Landau-level index and the longitudinal momentum. The results for the conductivity are presented in Sec.~\ref{QEDresults}. We study the dependence of the transverse and longitudinal conductivities on the temperature, the magnetic field, and the electrical chemical potential. In addition to QED plasma made of massive charge carriers, we also consider the chiral limit of QED plasma. In Sec.~\ref{QGPresults}, we extend the study to a strongly coupled QGP. Finally, we summarize our main results and conclusions in Sec.~\ref{Summary}. Several technical derivations and auxiliary results are given in the Appendixes at the end of the paper.

 \section{Electrical Conductivity}
 \label{sec:ElectricalConductivity}

 Using Kubo's linear response theory, one can express the electrical conductivity tensor $\sigma_{ij}$ in terms of the retarded current-current correlation function, 
 \begin{equation}
  \sigma_{ij}  = \lim_{\Omega\to 0}\frac{\mbox{Im}\left[\Pi_{ij}(\Omega+i0 ;\bm{0}) \right]}{\Omega}.
  \label{sigma-tensor}
 \end{equation}
 In a QED-like theory with $N_f$ fermion species having electric charges $e_f$, the correlation function (polarization tensor) at the leading one-loop order is given by the following expression \cite{Wang:2021ebh}:
 \begin{eqnarray}
  \Pi_{ij}(i\Omega_l;\bm{0}) &=&  4\pi \alpha \sum_{f=1}^{N_f} q_f^2 T\sum_{k=-\infty}^{\infty} 
  \int \frac{d^3 \bm{k}}{(2\pi)^3} \nonumber\\
  &\times&\mbox{tr} \left[ \gamma^i \bar{G}^f(i\omega_k ,\bm{k})  
  \gamma^j \bar{G}^f(i\omega_k-i\Omega_l,\bm{k})\right],
  \label{Pi-tensor}
 \end{eqnarray}
 where $\alpha =e^2/(4\pi)=1/137$ is the fine structure constant, $q_f = e_f/e$, and $e$ is the absolute value of the electron charge. By definition, $\bm{k}=(\bm{k}_\perp,k_z)$, where $\bm{k}_\perp=(k_x,k_y)$ and $k_z$ are the transverse and longitudinal momenta, respectively. Note, however, that the transverse part $\bm{k}_\perp$ does not represent  conventional momenta of any quantum states in a magnetized plasma. It is a momentum-like quantity introduced by the Fourier transform of the translation invariant part of the fermion propagator $\bar{G}^f(t-t^\prime,\bm{x}-\bm{x}^\prime)$. At the tree level, an explicit expression of the corresponding Fourier transform reads as  \cite{Miransky:2015ava}
 \begin{equation}
  \bar{G}^{f}(k_{0},\bm{k}) = ie^{-k_{\perp}^2 \ell_{f}^{2}}\sum_{n=0}^{\infty}
  (-1)^n \frac{D^{(0)}_{n,f}(k_{0},\bm{k})}{k_0^2-E_{f,n}^2},
  \label{free-prop}
 \end{equation}
 where $\ell_{f} = 1/\sqrt{|e_f B|}$ is the magnetic length,  $E_{f,n}=\sqrt{2n|e_f B|+m_f^2+k_z^2}$ are the Landau-level energies,  and
 \begin{eqnarray}
  D^{(0)}_{n,f}(k_{0},\bm{k}) &=& 2\left[(k_0\gamma_0-k_z\gamma^3 +m_f\right]\nonumber\\
   &\times&\big[{\cal P}_{+}L_n\left(2 k_{\perp}^2 \ell_{f}^{2}\right)-{\cal P}_{-}L_{n-1}\left(2 k_{\perp}^2 \ell_{f}^{2}\right)\big]
\nonumber\\
  &+& 4 (\bm{k}_{\perp}\cdot\bm{\gamma}_{\perp}) L_{n-1}^1\left(2 k_{\perp}^2 \ell_{f}^{2}\right).
  \label{Dn0-free}  
 \end{eqnarray}
 
 It is instructive to emphasize that we use a bare vertex in the expression for the polarization tensor in Eq.~(\ref{Pi-tensor}). This approximation is justified at the leading order in coupling. As demonstrated in Ref.~\cite{Hattori:2016cnt}, vertex corrections give only subleading contributions to the conductivity in strong magnetic fields. This observation is also consistent with findings in Ref.~\cite{Aarts:2002tn} concerning plasmas without magnetic fields.
 
 In Eq.~(\ref{Pi-tensor}), the effects of nonzero temperature are captured by using the imaginary-time formalism. By definition, the fermionic and bosonic Matsubara frequencies are $\omega_k= (2k+1)\pi T$ and $\Omega_l=2l \pi T$, respectively. After calculating the Matsubara sum, the retarded expression for the polarization tensor is obtained by performing the analytic continuation: $i\Omega_l\to \Omega+ i \epsilon$. To streamline the calculation, it is convenient to render the fermion propagator in the spectral form
 \begin{equation}
  \bar{G}^f(i\omega_k ,\bm{k})  = \int_{-\infty}^{\infty} \frac{dk_{0} A^f_{\bm{k}} (k_0) }{i\omega_k-k_{0}+\mu_f},
  \label{prop-spectral-fun}
 \end{equation}
 where $A^f_{\bm{k}} (k_0)$ is the spectral function and $\mu_f$ is the electrical chemical potential for fermions of flavor $f$. The spectral function $A^f_{\bm{k}} (k_0)$, which carries information about the quasiparticle properties in plasma, is formally defined in terms of the advanced and retarded propagators as follows:
 \begin{eqnarray}
  A^f_{\bm{k}} (k_0) =\frac{1}{2\pi i}\left[\bar{G}^f(k_0-i0,\bm{k}) -\bar{G}^f(k_0+i0,\bm{k})\right].
  \label{spectral-density-def}
 \end{eqnarray}
 After substituting the spectral representation Eq.~(\ref{prop-spectral-fun}) into Eq.~(\ref{Pi-tensor}), calculating the Matsubara sum, and performing the appropriate analytic continuation, we derive the following result for the imaginary part of the polarization tensor
  \begin{widetext}
    \begin{eqnarray}
\mbox{Im}\left[\Pi_{ij}(\Omega;\bm{0})\right]&=& -\frac{\alpha}{4\pi}\sum_{f=1}^{N_f} q_f^2 \int  dk_0 d^3 \bm{k} 
  \mbox{tr} \left[ \gamma^i A^f_{\bm{k}} (k_0) \gamma^jA^f_{\bm{k}} (k_0-\Omega) \right]
  \left( \tanh\frac{k_0-\mu_f}{2T}-\tanh\frac{k_0-\Omega-\mu_f}{2T}\right).
  \label{Im-Pi-tensor}
 \end{eqnarray}
   Then, substituting this expression into the definition for the conductivity tensor (\ref{sigma-tensor}), we finally obtain
 \begin{equation}
  \sigma_{ij}  =  -\frac{\alpha}{8\pi T}\sum_{f=1}^{N_f} q_f^2 \int  \frac{dk_0 d^3 \bm{k}}{\cosh^2\frac{k_0-\mu_f}{2T}}
  \mbox{tr} \left[ \gamma^i A^f_{\bm{k}} (k_0) \gamma^jA^f_{\bm{k}} (k_0) \right].
  \label{sigma-tensor-spectral}
 \end{equation}
  \end{widetext}
As is clear, the spectral density $A^f_{\bm{k}} (k_0)$ on the right-hand side encapsulates the main information that determines charge transport. Specifically, it depends on the damping rates $\Gamma_{f,n}(k_z)$ of charge carriers in their quantized Landau-level states \cite{Ghosh:2024hbf}, which are crucial for quantifying their individual contributions to plasma conductivity. 
 
 In a noninteracting theory, for example, the damping rate is zero, and the spectral density is represented by a sum of zero-width $\delta$-function peaks located at the exact particle energies, i.e., $k_{0}= \lambda E_{f,n}$ where $\lambda=\pm 1$. Consequently, each pair of coinciding peaks from the two spectral densities in Eq.~(\ref{sigma-tensor-spectral}) gives an infinite contribution to the conductivity. As we demonstrate later, this explains why the longitudinal conductivity ($\sigma_{\parallel} \equiv \sigma_{33}$) is indeed infinite in the free theory. In contrast, the transverse conductivity ($ \sigma_{\perp} \equiv \sigma_{11} =\sigma_{22} $) vanishes in the free theory. This seemingly surprising result arises from the Dirac structure of the spectral density in Eq.~(\ref{spectral-density-def}). For the transverse conductivity, the trace in Eq.~(\ref{sigma-tensor-spectral}) is proportional to products of spectral peaks from neighboring Landau levels that do not overlap. From a physics viewpoint, the magnetic field confines charges to fixed Landau orbits, thus preventing transport in the perpendicular directions. Only interactions allow these charges to escape trapping through transitions between neighboring Landau orbits, thereby contributing to the conductivity.
 
 In an interacting theory, the peaks in the spectral density have finite widths, which are determined by the particle damping rates. At weak coupling, an explicit expression for  $A^f_{\bm{k}} (k_0)$ can be derived from its free-theory counterpart by replacing the $\delta$-function peaks with Lorentzian functions of width $\Gamma_{f,n}(k_z)$,
 \begin{eqnarray}
  \delta(k_0-\lambda E_{f,n})\rightarrow \frac{1}{\pi}\frac{\Gamma_{f,n}}{\left(k_{0} -\lambda E_{f,n} \right)^2+\Gamma_{f,n}^2}.
 \end{eqnarray}
  Then, by making use of the fermion propagator in the Landau-level representation \cite{Miransky:2015ava} and using the definition in Eq.~(\ref{spectral-density-def}), we derive
 \begin{eqnarray}
  A^f_{\bm{k}} (k_0) &=& \frac{ie^{-k_\perp^2\ell_{f}^2}}{\pi}\sum_{\lambda=\pm}\sum_{n=0}^{\infty} 
  \frac{(-1)^n}{E_{f,n}}\Big\{
  \left[E_{f,n} \gamma^{0} 
  -\lambda  k_{z}\gamma^3+\lambda  m_f \right]\nonumber\\
  &\times&
  \left[{\cal P}_{+}L_n\left(2 k_\perp^2\ell_{f}^2\right)
  -{\cal P}_{-}L_{n-1}\left(2 k_\perp^2\ell_{f}^2\right)\right] \nonumber\\
  &+&2\lambda  (\bm{k}_\perp\cdot\bm{\gamma}_\perp) L_{n-1}^1\left(2 k_\perp^2 \ell_{f}^2\right)
  \Big\}\frac{\Gamma_{f,n}}{\left(k_{0} -\lambda E_{f,n} \right)^2+\Gamma_{f,n}^2} ,\nonumber\\
  \label{spectral-density}
 \end{eqnarray}
 where ${\cal P}_{\pm}=(1\pm i s_\perp \gamma^1\gamma^2)/2$ are the spin projectors, $s_\perp = \mbox{sign}(e_f B)$, and $L_n^{\alpha}\left(z\right)$ are the generalized Laguerre polynomials \cite{Gradshteyn:1943cpj}. To simplify the notation, an explicit dependence of $E_{f,n}$ and $\Gamma_{f,n}$ on the longitudinal momentum $k_z$ was suppressed. 
 
 Substituting spectral density (\ref{spectral-density}) into Eq.~(\ref{sigma-tensor-spectral}), calculating the traces, and integrating over $\bm{k}_\perp$, we derive the following expressions for the transverse and longitudinal conductivities,  
  \begin{widetext}
 \begin{eqnarray}
  \sigma_{\perp} &=&
  \frac{\alpha}{\pi^2T}\sum_{f=1}^{N_f} \frac{q_f^2}{\ell_f^2 }
  \sum_{n=0}^{\infty}
  \int_{-\infty}^{\infty} \int_{-\infty}^{\infty} \frac{d k_{0} d k_{z} }{\cosh^2\frac{k_{0}-\mu_f}{2T}}   \frac{\Gamma_{f,n+1} \Gamma_{f,n}
   \left[ \left(k_{0}^2+E_{f,n}^2+\Gamma_{f,n}^2\right)\left(k_{0}^2+E_{f,n+1}^2+\Gamma_{f,n+1}^2\right)  -4k_{0}^2\left(k_z^2 +m_f^2\right)\right]}
  {\left[\left(E_{f,n}^2+\Gamma_{f,n}^2 -k_{0}^2\right)^2+4k_{0}^2 \Gamma_{f,n}^2 \right]
   \left[\left(E_{f,n+1}^2+\Gamma_{f,n+1}^2 -k_{0}^2\right)^2+4k_{0}^2 \Gamma_{f,n+1}^2 \right]},
  \label{conductivity-perp}
\\
  \sigma_{\parallel} &=& \frac{\alpha}{2\pi^2 T}\sum_{f=1}^{N_f}  \frac{q_f^2}{\ell_f^2 }
  \sum_{n=0}^{\infty}\beta_n
  \int_{-\infty}^{\infty} \int_{-\infty}^{\infty} \frac{d k_{0} d k_{z} }{\cosh^2\frac{k_{0}-\mu_f}{2T}}
  \frac{\Gamma_{f,n}^2\left[\left(E_{f,n}^2+\Gamma_{f,n}^2-k_{0}^2\right)^2 +4k_{0}^2\left(2k_z^2+\Gamma_{f,n}^2\right)\right]}
  {\left[\left(E_{f,n}^2+\Gamma_{f,n}^2 -k_{0}^2\right)^2+4k_{0}^2 \Gamma_{f,n}^2 \right]^2} , 
  \label{conductivity-parallel}
 \end{eqnarray}
  \end{widetext}
   respectively.  In the derivation, we used the results for Dirac traces and two types of sums in Appendix~\ref{app:traces}. In the expression for  longitudinal conductivity, we introduced the shorthand notation $\beta_n\equiv 2-\delta_{n,0}$ to represent the spin degeneracy factor. These general expressions for the conductivities are our main analytical results. 
 
 Note that the Hall conductivity $\sigma_{\rm H}  \equiv \sigma_{12} = - \sigma_{21}$ is also nonvanishing when the chemical potential $\mu_f$ is nonzero. Moreover, its leading contribution is of zeroth order in coupling and is proportional to the electric charge density. The Hall conductivity is an example of nondissipative transport, which explains why it remains nonzero even in the free theory. In this study, we primarily focus on the dissipative transport represented by the transverse and longitudinal conductivities.
 
 In the limit of a small damping rate, assuming that the spectral peaks from neighboring Landau levels do not overlap and are well separated, the integration over $k_0$ in Eqs.~(\ref{conductivity-perp}) and (\ref{conductivity-parallel}) can be performed approximately. Before deriving the corresponding approximate results, let us note that the expressions in the square brackets in the denominators of Eqs.~(\ref{conductivity-perp}) and (\ref{conductivity-parallel}) can be further factorized, e.g., 
  \begin{widetext}
  \begin{eqnarray}
  \left(E_{f,n}^2+\Gamma_{f,n}^2 -k_{0}^2\right)^2+4k_{0}^2 \Gamma_{f,n}^2  &=& \left[\left(k_0-E_{f,n}\right)^2+\Gamma_{f,n}^2 \right] \left[\left(k_0+E_{f,n}\right)^2+\Gamma_{f,n}^2 \right].
 \end{eqnarray}
 By using such factorizations and rendering the integrands in the form of partial fractions, one can then reduce the energy integrations to the following two types:
 \begin{eqnarray}
  \frac{1}{\pi} \int_{-\infty}^{\infty}  \frac{ dk_0 F\left(k_0\right) \Gamma_1^2}{\left[\left(k_{0} -E_{1} \right)^2+\Gamma_1^2\right]^2} &=&  \frac{F\left(E_1\right)}{2\Gamma_1} ,
  \\
  \frac{1}{\pi} \int_{-\infty}^{\infty}  \frac{ dk_0 \Gamma_1 \Gamma_2 F\left(k_0\right)  }
  {\left[\left(k_{0} - E_{1} \right)^2+\Gamma_1^2\right] \left[\left(k_{0} - E_{2} \right)^2+\Gamma_{2}^2\right]}
  &=& \frac{\Gamma_2  F\left(E_1\right)}{\left(E_{1} - E_{2} \right)^2+\Gamma_{2}^2} 
  + \frac{\Gamma_1  F\left(E_2\right)}{\left(E_{1} - E_{2} \right)^2+\Gamma_{1}^2},
 \end{eqnarray}
 where $F\left(k_0\right)$ is a slowly varying function near $k_0=E_{1}$ and $k_0=E_{2}$. 
 
 Thus, in the limit of narrow spectral peaks, from Eqs.~(\ref{conductivity-perp}) and (\ref{conductivity-parallel}), we derive the following approximate expressions for the transverse and longitudinal conductivities 
 \begin{eqnarray}
  \sigma_{\perp} &\approx & \frac{\alpha}{2\pi T}\sum_{f=1}^{N_f} \frac{q_f^2}{\ell_f^2 }
  \sum_{n=0}^{\infty}
  \int_{0}^{\infty}  d k_{z}\left(\frac{1}{\cosh^2\frac{E_{f,n}-\mu_f}{2T}} + \frac{1}{\cosh^2\frac{E_{f,n}+\mu_f}{2T}} \right)  \frac{ \Gamma_{f,n+1} f\left(E_{f,n}\right)}{\left[  (E_{f,n}-E_{f,n+1})^2+ \Gamma_{f,n+1}^2\right]
   \left[   (E_{f,n}+E_{f,n+1})^2+ \Gamma_{f,n+1}^2 \right] }\nonumber\\
  &+&\frac{\alpha}{2\pi T}\sum_{f=1}^{N_f} \frac{q_f^2}{\ell_f^2 }
  \sum_{n=0}^{\infty}
  \int_{0}^{\infty} d k_{z} \left(\frac{1}{\cosh^2\frac{E_{f,n+1}-\mu_f}{2T}} + \frac{1}{\cosh^2\frac{E_{f,n+1}+\mu_f}{2T}} \right) 
  \frac{ \Gamma_{f,n} f\left(E_{f,n+1}\right)}{\left[  (E_{f,n}-E_{f,n+1})^2+ \Gamma_{f,n}^2\right]
   \left[   (E_{f,n}+E_{f,n+1})^2+ \Gamma_{f,n}^2 \right] },  \nonumber\\
  \label{conductivity-perp-approx}
\\
  \sigma_{\parallel} &\approx & \frac{\alpha}{4\pi T} \sum_{f=1}^{N_f} \frac{q_f^2}{\ell_f^2 }
  \sum_{n=0}^{\infty}\beta_n
  \int_{0}^{\infty} \frac{d k_{z}  \left(2 k_{z}^2+ \Gamma_{f,n}^2\right) }
  {\Gamma_{f,n} \left(2  E_{f,n}^2+ \Gamma_{f,n}^2 \right)  } \left( \frac{1}{\cosh^2\frac{E_{f,n}-\mu_f}{2T}}+\frac{1}{\cosh^2\frac{E_{f,n}+\mu_f}{2T}}
  \right),
  \label{conductivity-parallel-approx}
 \end{eqnarray}
   \end{widetext}
where we introduced the shorthand notation
 \begin{eqnarray}
  f(k_0) &=& \frac{\left(k_{0}^2+E_{f,n}^2+\Gamma_{f,n}^2\right)\left(k_{0}^2+E_{f,n+1}^2+\Gamma_{f,n+1}^2\right)}{k_0^2} \nonumber\\
  &-&4 \left(k_z^2 +m_f^2\right).
  \label{f-k0}
 \end{eqnarray}
  
 It should be emphasized that the approximate expressions for the conductivities in Eqs.~(\ref{conductivity-perp-approx}) and (\ref{conductivity-parallel-approx}) should be used with great caution. Their validity is limited to the weakly coupled regime, where the damping rates are sufficiently small to prevent any overlaps of spectral peaks of different Landau-level states. Additionally, by scrutinizing the expression the transverse conductivity, which is given in terms of function $f(k_0)$ in Eq.~(\ref{f-k0}), one finds that the approximation fails in the massless limit. Indeed, the lowest Landau-level energy (with $k_0^2=k_z^2$ at $m_f=0$) gives a divergent contribution due to a singularity in the integrand at the vanishing longitudinal momentum. This is clearly an artifact of the approximation, as the original expression (\ref{conductivity-perp}) is finite even in the massless limit. The same reasoning extends to the case of superstrong magnetic fields, even if the fermion mass is nonzero but $m_f^2/|e_f B|$ is small. In such a regime, the approximate result for the transverse conductivity becomes unreliable due to an artificially increased sensitivity of the integrand to the infrared region near $k_z=0$. 
 
 Despite their limited validity, the approximate expressions for the transverse and longitudinal conductivities in Eqs.~(\ref{conductivity-perp-approx}) and (\ref{conductivity-parallel-approx}) are invaluable for providing a deeper insight into the underlying mechanism of charge transport in strongly magnetized plasmas. As one can see, both conductivities are given by weighted sums of Landau-level contributions. The weights are determined by the plasma temperature $T$, the electric chemical potentials of individual fermion flavors $\mu_f$, as well as the damping rates $\Gamma_{f,n}$. 
 
 As we see from Eq.~(\ref{conductivity-parallel-approx}), different Landau levels contribute to the longitudinal conductivity as distinct species of charge carriers. The explicit expression suggests that each level contributes proportionally to the inverse damping rate: $\sigma_{\parallel} = \sum_{f,n} \sigma_{\parallel,f,n}$, where $\sigma_{\parallel,f,n} \propto 1/\Gamma_{f,n}$. In the free theory (with $\Gamma_{f,n} \to 0$), in particular, the longitudinal conductivity becomes infinite. 
 
 In contrast, the partial contributions to the transverse conductivity in Eq.~(\ref{conductivity-perp-approx}) are associated with quantum transitions between adjacent Landau levels with indices $n$ and $n+1$. Furthermore, the individual contributions are proportional to the damping rates rather than their inverse values, i.e., $\sigma_{\perp,f,n} \propto \Gamma_{f,n}$. In the free theory (with $\Gamma_{f,n} \to 0$), therefore, the transverse conductivity of a strongly magnetized plasma vanishes. This indicates that the underlying mechanism for transverse transport is qualitatively different. This difference is not surprising, as a strong background magnetic field confines charged carriers in the transverse direction. As mentioned earlier, conduction occurs only as a result of interactions that provide a transport pathway via quantum transitions between Landau levels \cite{Ghosh:2024fkg}.
 
 \section{Damping rate}
 \label{sec-Damping-rate}
 
 To calculate the electrical conductivities given by Eqs.~(\ref{conductivity-perp}) and (\ref{conductivity-parallel}), or their approximate counterparts in Eqs.~(\ref{conductivity-perp-approx}) and (\ref{conductivity-parallel-approx}), one needs to know the damping rates $\Gamma_{f,n}(k_z)$ for the charge carriers in a strongly magnetized plasma. Using the Landau-level representation, the corresponding results were obtained in Ref.~\cite{Ghosh:2024hbf} within a gauge theory framework at the leading order in coupling. 
 
 As shown in Ref.~\cite{Ghosh:2024hbf}, the damping rate $\Gamma_{f,n}(k_z)$ of a state with a given Landau-level index $n$ is determined by the three types of quantum processes: (i) transitions to Landau levels with lower indices $n^\prime$  ($\psi_{n}\to \psi_{n^\prime} +\gamma$ with $n>n^\prime$), (ii) transitions to Landau levels with higher indices $n^\prime$ ($\psi_{n}+\gamma\to \psi_{n^\prime} $ with $n<n^\prime$), and (iii) transitions to Landau-level states with negative energies (i.e., the annihilation process $\psi_{n}+\bar{\psi}_{n^\prime}\to \gamma$ for any $n$ and $n^\prime$).  
 
 The leading-order contributions are of the order of $\alpha |e_{f}B|/T$. In the absence of a magnetic field, however, the corresponding one-to-two and two-to-one processes are kinematically forbidden. Instead, the damping rate is determined by the two-to-two processes, i.e., $\psi_{n}+\gamma\to \psi_{n^\prime}+\gamma$ and $\psi_{n}+\bar{\psi}_{n^\prime}\to \gamma+\gamma$, which contribute at order $\alpha^2 T$. In the rest of this study, therefore, we will assume that $|e_{f}B|/T^2\gg \alpha$, which guarantees that the leading-order processes dominate. 
 
 The expression for the spin-averaged damping rate of a particle state with Landau-level index $n$ reads as \cite{Ghosh:2024hbf} 
   \begin{widetext}  
   \begin{equation}
  \Gamma^{\rm (ave)}_{f,n}(k_z) = \frac{\alpha_{*}}{2 \beta_n  \ell_f^2 E_{f,n}} 
  \sum_{n^{\prime}=0}^\infty   \sum_{s^\prime=\pm 1}   \sum_{s_1=\pm 1}  \sum_{s_2=\pm 1} 
  \int   d\xi   \frac{{\cal M}_{n,n^{\prime}} (\xi) \left[ 1-n_F(s_1 E_{f,n^{\prime},s^\prime})+n_B(s_2 E_{q,s^\prime}) \right] }{s_1 s_2 \sqrt{(\xi-\xi^{-})(\xi-\xi^{+})} } , 
  \label{Gamma_n_pz-short}
 \end{equation}
   \end{widetext}
where $\xi^{\pm}=\frac{1}{2}\left[\sqrt{2n^{\prime}+(m_{f}\ell_f)^2}\pm \sqrt{2n+(m_{f}\ell_f)^2}\right]^2$ are the threshold parameters that depend on both Landau-level indices $n$ and $n^{\prime}$. For simplicity, we omit these indices in the notation to avoid making analytical expressions unnecessarily cumbersome.

We would like to stress that Eq.~(\ref{Gamma_n_pz-short}) defines the spin-averaged damping rate for charged particles in the $n$th Landau level. This rate was obtained by generalizing Weldon's well-known method \cite{Weldon:1983jn}, which was previously applied only to cases with vanishing magnetic fields, to a strongly magnetized plasma in Ref.~\cite{Ghosh:2024hbf}. Since the lowest Landau level contains only one spin state, while Weldon's method inherently accounts for two spin states, we corrected the expression by introducing the Landau-level dependent spin degeneracy factor $\beta_n\equiv 2-\delta_{n,0}$ in the denominator of Eq.~(\ref{Gamma_n_pz-short}). This adjustment ensures that the damping rates in both the lowest and higher Landau levels are accurate and consistent with the results obtained from the pole locations in the fermion propagator \cite{Ghosh:2024hbf}.
 
In the rate definition in Eq.~(\ref{Gamma_n_pz-short}), the coupling constant $\alpha_{*}$ is given by $\alpha q_f^2$ in a QED-like theory, where fermions have flavor-dependent electric charges $e_f =q_f e$. In QCD, on the other hand, $\alpha_{*}=\alpha_{s} C_F$ where $ \alpha_{s} = g^2/(4\pi)$ is the strong coupling constant and $C_F=(N_c^2-1)/(2N_c)=4/3$, assuming $N_c=3$. 

 The energies of the final fermion and the photon that appear in the distribution functions in Eq.~(\ref{Gamma_n_pz-short}) are given by the following expressions:
 \begin{eqnarray}
  s_1 E_{f,n^{\prime},s^\prime}  &=& 
  \frac{E_{f,n}}{2}\left(1+\frac{2n^{\prime}+m_{f}^2\ell_f^2-2\xi}{2n +m_{f}^2\ell_f^2}  \right)
  \nonumber\\
&&  + s^\prime k_z \frac{\sqrt{(\xi-\xi^{-})(\xi-\xi^{+})}}{ 2n+m_{f}^2\ell_f^2} ,
  \label{Enkz} \\
  s_2 E_{q,s^\prime} &=& 
  \frac{E_{f,n}}{2}\left(1-\frac{2n^{\prime}+m_{f}^2\ell_f^2-2\xi}{2n +m_{f}^2\ell_f^2}  \right)
    \nonumber\\
&&- s^\prime k_z \frac{\sqrt{(\xi-\xi^{-})(\xi-\xi^{+})}}{ 2n+m_{f}^2\ell_f^2} ,
  \label{Eqkz}
 \end{eqnarray}
 respectively. These are obtained by explicitly solving the energy conservation equation for the underlying one-to-two and two-to-one processes. It should be noted that, for simplicity, we retain only the original indices in the notation for the fermion and photon energies, $E_{f,n^{\prime},s^\prime}$ and $E_{q,s^\prime}$, respectively, even though they depend on both Landau level indices  $n$ and $n^{\prime}$, as well as other kinematic parameters.
 
Function ${\cal M}_{n,n^{\prime}} (\xi)$ in the integrand of Eq.~(\ref{Gamma_n_pz-short}) is determined by the squared amplitude of the leading-order processes \footnote{The proof that the squared amplitude of the leading-order processes is proportional to function ${\cal M}_{n,n^{\prime}} (\xi)$ is given in Appendix~\ref{app:amplitude}.}.
 Its explicit form is given by
 \begin{eqnarray}
{\cal M}_{n,n^{\prime}}(\xi)  &=&-  \left(n+n^{\prime}+ m_{f}^2\ell_f^2\right)\left[\mathcal{I}_{0}^{n,n^{\prime}}(\xi)+\mathcal{I}_{0}^{n-1,n^{\prime}-1}(\xi) \right]  \nonumber\\
&&
  +(n+n^{\prime}) \left[\mathcal{I}_{0}^{n,n^{\prime}-1}(\xi)+\mathcal{I}_{0}^{n-1,n^{\prime}}(\xi) \right],
  \label{Mnnp}
 \end{eqnarray} 
and function $\mathcal{I}_{0}^{n,n^{\prime}}(\xi)$ is defined in Eq.~(\ref{I0f-LL-form1}) in Appendix~\ref{app:amplitude}.
 Note that the integration range in $\xi$ depends on the choice of signs $s_1$ and $s_2$, representing different types of processes, i.e.,
 \begin{eqnarray}
  \psi_{n}\to \psi_{n^\prime} +\gamma \quad (s_1>0, ~ s_2>0): &\quad& 0<\xi<\xi^{-},
  \label{proc-1}\\
  \psi_{n}+\gamma\to \psi_{n^\prime} \quad (s_1>0, ~ s_2<0): &\quad& 0<\xi<\xi^{-},
  \label{proc-2}\\
  \psi_{n}+\bar{\psi}_{n^\prime}\to \gamma \quad (s_1<0, ~ s_2>0): &\quad& \xi^{+}<\xi< \infty. 
  \label{proc-3}
 \end{eqnarray}
 In the next two sections, we use the spin-averaged damping rate $\Gamma^{\rm (ave)}_{f,n}(k_z)$ in order to calculate the electrical conductivities, defined by Eqs.~(\ref{conductivity-perp}) and (\ref{conductivity-parallel}), for QED and QCD plasmas. Before proceeding to those calculations, several comments are in order. 
 
Firstly, one may wonder whether it would be preferable to use a more refined approximation for the rate without relying on spin averaging. The corresponding rates for the spin-up and spin-down states, $\Gamma_{f,n}^{(+)}$ and $\Gamma_{f,n}^{(-)}$, in higher Landau levels ($n\geq1$) were obtained in Ref.~\cite{Ghosh:2024hbf}. They were extracted from the location of poles in the full propagator. It was found, however, that the rates for both spin states are very close to the spin-averaged value, $\Gamma_{f,n}^{\rm (ave)} \equiv \left(\Gamma_{f,n}^{(+)}+\Gamma_{f,n}^{(-)}\right)/2$. 
 
The other comment concerns the chiral limit in the strongly magnetized plasma. By examining the analytic expression for the rate in  Eq.~(\ref{Gamma_n_pz-short}), one finds that the massless limit is singular. The problem stems from the contribution of the lowest Landau level. When one sets $m_f=0$ and $n=0$, the energies in Eqs.~(\ref{Enkz}) and (\ref{Eqkz}) become ill-defined. The singularities are an artifact of additional kinematic constraints on the one-to-two and two-to-one processes involving the gapless Landau level when $m_f=0$. To resolve the issue, below we rederive the rate in the lowest Landau level in the chiral limit. 
 
 By analyzing the energy-momentum conservation in the massless case, we find that the final fermion and the photon energies change when $n=0$. Specifically, the expressions in Eqs.~(\ref{Enkz}) and (\ref{Eqkz}) are replaced by
 \begin{eqnarray}
  s_1 E_{f,n^{\prime}}  &=& - \frac{(\xi -n^{\prime})^2 +2n^{\prime} k_z^2\ell_f^2 }{2\ell_f^2 |k_z|(\xi-n^{\prime})} , 
  \label{Enkz-chiral} \\
  s_2 E_{q}  &=&  \frac{|e_f B|(\xi -n^{\prime})^2 +2\xi k_z^2}{2 |k_z| (\xi -n^{\prime})} ,
  \label{Eqkz-chiral} 
 \end{eqnarray}
 respectively. Note that, unlike the $n=0$ case at nonzero mass, where there are two different solutions labeled by $s^\prime$, there is only one solution when $m_f=0$. After substituting these energies into Eq.~(\ref{Gamma_n_pz-short}), we calculate the corrected rate in the lowest Landau level ($n=0$). As for the damping rates in higher Landau levels ($n\geq 1$), their expressions remain the same but are evaluated at $m_f=0$.

 \section{Electrical conductivity in QED plasma}
 \label{QEDresults}
 
 Let us now apply the general theory of Secs.~\ref{sec:ElectricalConductivity} and \ref{sec-Damping-rate} to calculate the transverse and longitudinal electrical conductivities in a magnetized QED plasma, made of electrons and positrons. In this case, we have a single flavor ($N_f=1$) of charged carriers with $e_f=-e$ and $m_f=m_e =0.511~\mbox{MeV}$. Because of the smallness of the coupling constant in QED, the expected range of validity of the leading-order approximation, $|eB|\gtrsim \alpha T^2$, may extends down to moderately weak magnetic fields.  
 
\begin{figure}[b]
  \centering
  \includegraphics[width=0.98\columnwidth]{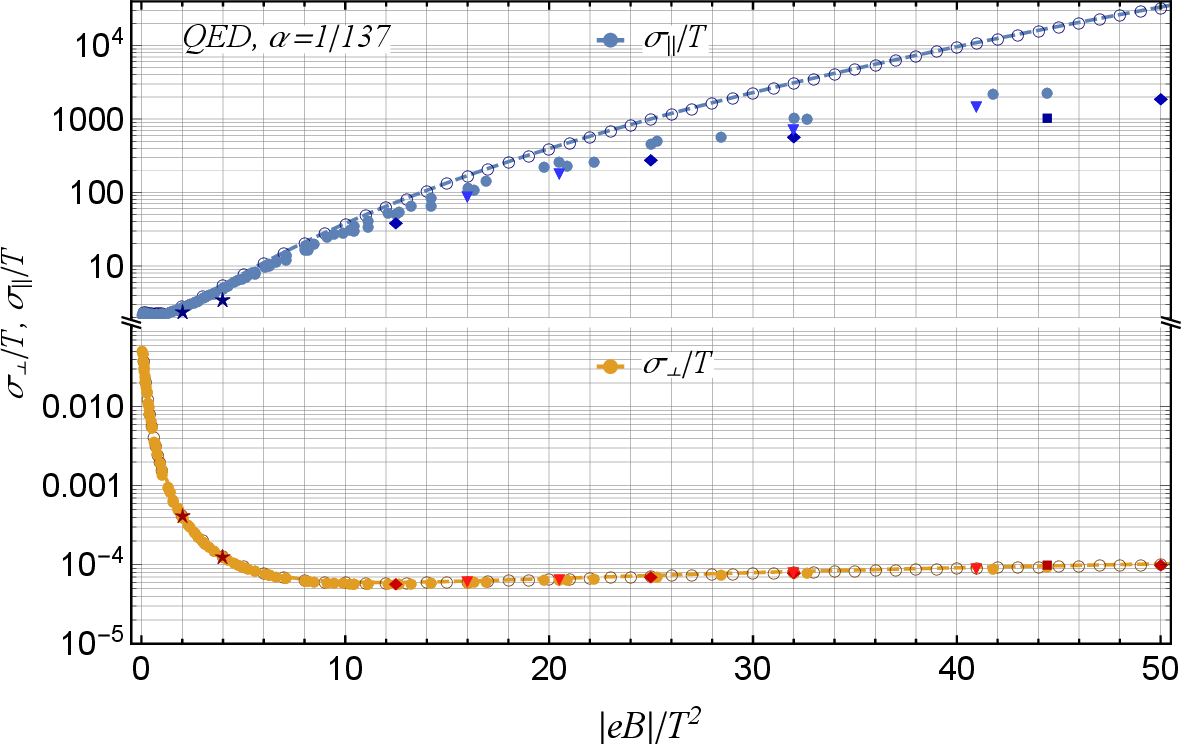} 
  \caption{The transverse and longitudinal conductivities in units of temperature as functions of the dimensionless ratio $|eB|/T^2$. Empty circles and interpolating dashed lines represent the conductivities in the chiral limit. Special symbols represents the results for the lowest temperatures: $T=2.5m_e$ (stars),  $T=15m_e$ (squares),  $T=20m_e$ (rhombi),  $T=25m_e$ (triangles).}
  \label{fig:cond_QED}
 \end{figure}

 Before calculating the conductivities, one needs to obtain the damping rates first. Thus, we start by producing tabulated data sets for the electron damping rates in a wide range of temperatures, $15 m_e \leq T \leq 80 m_e$, and magnetic fields, $(15m_e)^2 \leq |eB| \leq (200m_e)^2 $. We generate numerical data for up to $n_{\rm max}=50$ Landau levels and about $50$ points of the longitudinal momentum. When using Eq.~(\ref{Gamma_n_pz-short}) to calculate the rates, we account for all processes that include the final electron states with Landau-level indices up to $n_{\rm max}^\prime =2n_{\rm max}$. The corresponding data is then interpolated before used in numerical calculation of the conductivities defined by Eqs.~(\ref{conductivity-perp}) and (\ref{conductivity-parallel}). It should be mentioned that, in a wide range of model parameters, the transverse and longitudinal conductivities in QED can be well approximated by Eqs.~(\ref{conductivity-perp-approx}) and (\ref{conductivity-parallel-approx}). Nevertheless, all numerical data presented in Fig.~\ref{fig:cond_QED} below are obtained by using the exact expressions in Eqs.~(\ref{conductivity-perp}) and (\ref{conductivity-parallel}). 
 
 \subsection{Scaling dependence of dimensionless conductivities and the chiral limit}
 \label{subsec:Scaling-chiral}

 When considering sufficiently high temperatures ($T\gg m_e$) and strong magnetic fields ($|eB|\gg m_e^2$), we might expect that effects of a nonzero electron mass are negligible. Then, using dimensional arguments, one can argue that the conductivities measured in units of temperature, $\tilde{\sigma}_{\perp}\equiv \sigma_{\perp}/T$ and $\tilde{\sigma}_{\parallel}\equiv \sigma_{\parallel}/T$, should be well approximated by some universal functions of the dimensionless ratio $|eB|/T^2$. (At this point, we assume that the electron chemical potential is zero.) To some degree, the universal scaling is indeed supported by our numerical data for the conductivities in Fig.~\ref{fig:cond_QED}. 
 
 In Fig.~\ref{fig:cond_QED}, we present a compilation of over hundred data points for the transverse and longitudinal conductivities at various temperatures and magnetic fields.  The corresponding numerical values of the conductivities are provided as data files in the Supplemental Material \cite{QED-QCDCond:2024}. Filled markers represent the data for QED with a nonzero fermion mass. 
 
 For comparison, we also include results for a massless (chiral) QED plasma, represented by empty circles. In the figure, we use dashed lines as a convenient visual guide to interpolate the chiral data. The latter are well approximated by the following Pad\'{e} approximants:
 \begin{widetext}  
 \begin{eqnarray}
  \tilde{\sigma}_{\perp, {\rm chiral}}^{\rm QED} &\simeq & 0.076 \frac{1 + 0.13 b^{3/2} -0.002 b^2 + 0.0026 b^{7/2} }
  {1 + 6.5 b + 31.6 b^2 + 13.2 b^3 },
  \label{signa11QEDfit-chiral} 
\\
  \tilde{\sigma}_{\parallel, {\rm chiral}}^{\rm QED}  &\simeq&  
  2.47 \frac{ 1 - 0.172 b + 0.125 b^2 - 0.0058 b^3 + 8.1\times 10^{-4} b^4 - 8.6\times 10^{-6} b^5 + 2.88\times 10^{-7} b^6}
  {1 - 0.013 b + 5.4\times 10^{-5}b^2 - 5.6\times 10^{-8} b^3}  , 
  \label{signa33QEDfit-chiral}
 \end{eqnarray} 
  \end{widetext}  
  where $b=|eB|/T^2$. The quality of these fits are better than about $2\%$ in the whole range of values $0.1\lesssim b\lesssim 100$. It should be mentioned that $ \tilde{\sigma}_{\perp, {\rm chiral}}^{\rm QED} $ and $ \tilde{\sigma}_{\parallel, {\rm chiral}}^{\rm QED} $ have local minima at $b_{{\rm min},\perp}\approx 12$ and $b_{{\rm min},\parallel}\approx 0.6$, respectively.

 We observe from Fig.~\ref{fig:cond_QED} that the QED data closely approaches the chiral limit over a wide range of small to moderately larges values of $|eB|/T^2$. However, significant deviations occur when $|eB|/T^2\gtrsim 10$. In the latter region, the lowest Landau level is expected to provide the dominant contribution to charge transport. To further illuminate the effect, we used special symbols to mark the data points for several lowest temperatures, $T=2.5m_e$ (stars),  $T=15m_e$ (squares),  $T=20m_e$ (rhombi),  $T=25m_e$ (triangles), where $m_e$ is the electron mass. As seen, the smaller the values of $T/m_e$ or the larger the values of $|eB|/T^2$, the more significant deviations from the chiral limit. As discussed in the previous two sections, the expressions for the conductivities and the damping rates are highly sensitive to the kinematics of the lowest Landau level in the infrared region near $k_z=0$ when $m_e \to 0$. Therefore, it is not surprising that deviations from the chiral limit are substantial at low temperatures and strong magnetic fields.

 \begin{figure*}[t]
   \begin{center}
  \includegraphics[width=0.98\columnwidth]{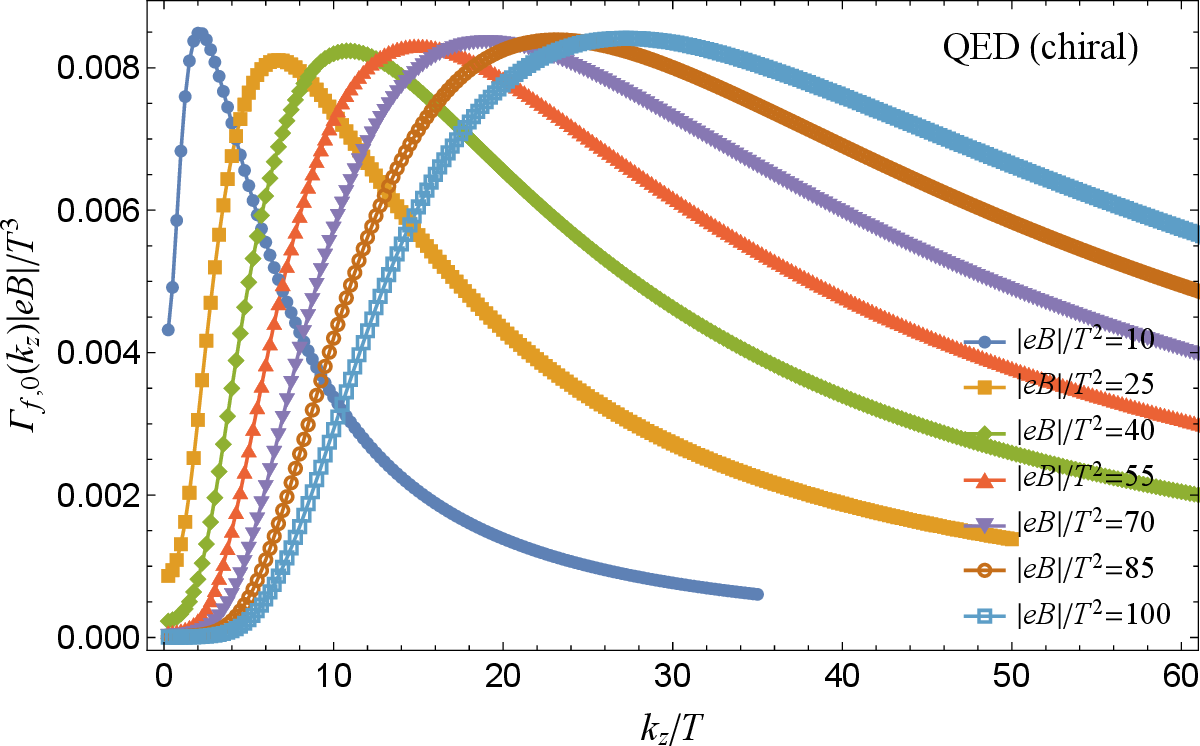} 
  \hspace{0.05\columnwidth}
  \includegraphics[width=0.98\columnwidth]{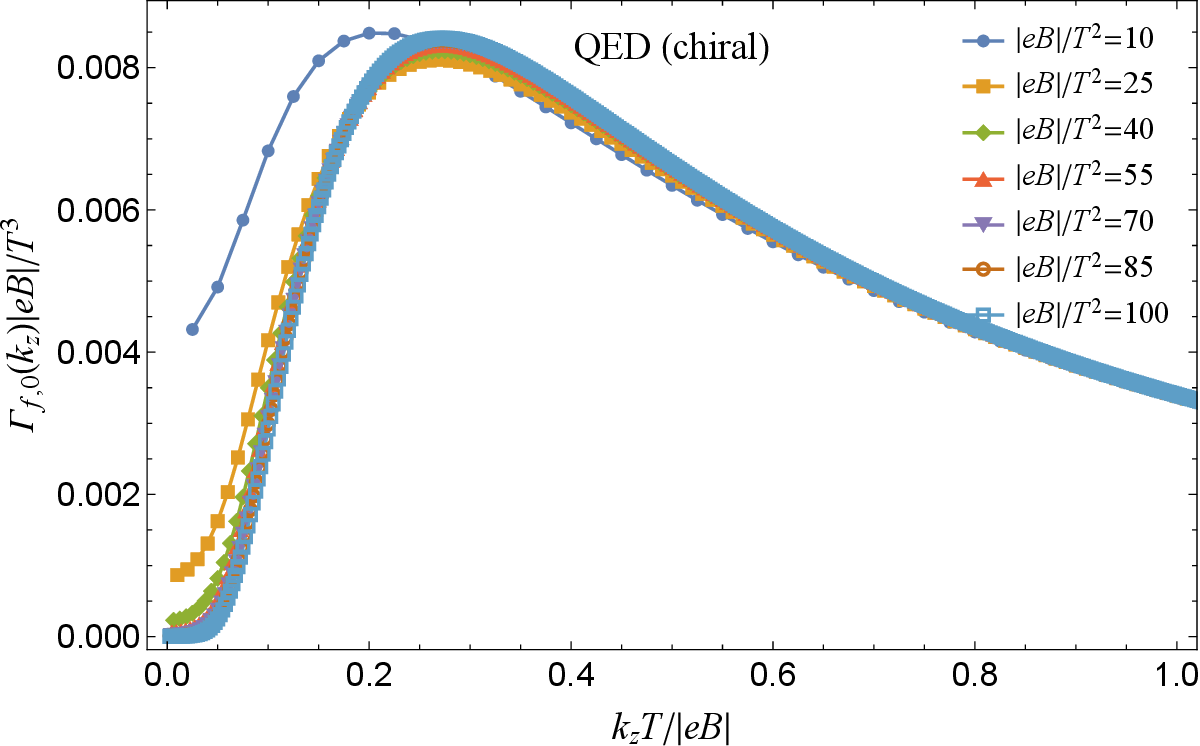} 
  \caption{Scaling behavior of the damping rate in the lowest Landau level for several fixed values of the magnetic field.}
  \label{fig:Gamma0-scaling}
  \end{center}
 \end{figure*}
 
 In the strong-field limit where $|eB|/T^2\gg 1$, longitudinal transport is dominated by charge carriers in the lowest Landau level. Naively, since the Landau-level  density of states is proportional to the field strength, one might expect that the longitudinal conductivity $\tilde{\sigma}_{\parallel}$ would also be proportional to the field. However, by analyzing the data in Fig.~\ref{fig:cond_QED}, we find that $\tilde{\sigma}_{\parallel}$ increases much more rapidly than linearly with the magnetic field when $|eB|/T^2 \gg 1$. In fact, the numerical data in the chiral limit indicates that the longitudinal conductivity might be growing as $\tilde{\sigma}_{\parallel}\propto b^\gamma$, where $b=|eB|/T^2$ and the power $\gamma$ might be as large as $6$ to $8$. (By comparison, in the case of nonzero mass, the growth appears to be also a power law but with a smaller value of $\gamma$.) This additional enhancement is attributed to the damping rate $\Gamma_{f,0}(k_z)$, which is suppressed by the magnetic field. 
 
To support the claim, we show representative results for the damping rates in the lowest Landau level at several fixed values of $|eB|/T^2$ in Fig.~\ref{fig:Gamma0-scaling}. For simplicity, here we consider the chiral limit, which involves fewer dimensionless parameters. To demonstrate that the maximum damping rate scales as $T^3/|eB|$, we present numerical results for the rescaled quantity $\Gamma_{f,0}(k_z)|eB|/T^3$ as a function of the longitudinal momentum. As shown in the left panel, the maximum values of the rescaled rates are approximately constant, although their dependence on $k_z$ differs. As the magnetic field increases, the peak positions shift to higher $k_z$ values. 

After rescaling the longitudinal momentum with the magnetic field, we replot the data in the right panel in Fig.~\ref{fig:Gamma0-scaling}. Observing that nearly all the rescaled data aligns along the same curve, we reconfirm that the maximum damping rates are inversely proportional to $|eB|$, and the locations of their maxima scale linearly with $|eB|$.

The damping rates for the lowest Landau level, presented in Fig.~\ref{fig:Gamma0-scaling}, are calculated by fully accounting for transitions to all higher Landau levels. Moreover, contrary to a naive expectation, transitions within the lowest Landau level alone do not dominate the numerical results for $\Gamma_{f,0}(k_z)$, even under very strong magnetic fields. This is especially important in the limit of a small (or vanishing) fermion mass or nonzero longitudinal momenta. For instance, our numerical results at $|eB|/T^2\simeq 50$ reveal that quantum transitions to the $n^\prime = 1$ and even $n^\prime = 2$ Landau levels play a significant role in determining $\Gamma_{f,0}(k_z)$. This is not unexpected, as the naive strong-field approximation, where only transitions within the lowest Landau level are considered, produces a zero damping rate when $m_f \to 0$ \cite{Hattori:2016cnt,Hattori:2016lqx}. Indeed, as seen from Eq.~(\ref{Mnnp}), the corresponding squared amplitude ${\cal M}_{0,0}(\xi)$ is proportional to $ m_f^2$, which vanishes in the chiral limit. Therefore, the primary contributions to $\Gamma_{f,0}(k_z)$ come from transitions to higher Landau levels when the fermion mass is small or vanishing. One can also verify that transitions to higher Landau levels dominate at sufficiently large longitudinal momenta $k_z$ even when mass is not very small.
 
 Let us now turn to the behavior of the transverse conductivity $\tilde{\sigma}_{\perp}$ at large $b=|eB|/T^2$. As shown in Fig.~\ref{fig:cond_QED}, it also increases as $b$ becomes sufficiently large, albeit much more slowly than $\tilde{\sigma}_{\parallel}$. Numerically, its functional dependence is well approximated by $\tilde{\sigma}_{\perp}\simeq 1.5\times 10^{-5} \sqrt{b}$, which is an outcome of two competing effects: the growing degeneracy of states and the suppression of damping rates in the $n=0$ and $n=1$ Landau levels. Ultimately, the increase in degeneracy slightly outweighs the suppression of the damping rates.
 
 In general, when the fermion mass is nonzero, the dimensionless conductivities in QED, $\tilde{\sigma}_{\perp}$ and $\tilde{\sigma}_{\parallel}$, should be given by some functions that depend on the two unitless ratios, $|eB|/T^2$ and $m_e/T$. Therefore, in the utrarelativistic regime, which assumes $|eB|\gg m_e^2$ and $T\gg m_e$, one might expect that the results can be approximated by the following Taylor expansions: 
 \begin{eqnarray}
  \tilde{\sigma}_{\perp} &\simeq& \tilde{\sigma}_{\perp, {\rm chiral}}\left(b\right) + \frac{m_e}{T} f_{\perp,1}\left(b\right) + \frac{m_e^2}{T^2} f_{\perp,2}\left(b\right)  + \cdots , \\
  \tilde{\sigma}_{\parallel} &\simeq & \tilde{\sigma}_{\parallel, {\rm chiral}}\left(b\right)  + \frac{m_e}{T} f_{\parallel,1}\left(b\right) + \frac{m_e^2}{T^2} f_{\parallel,2}\left(b\right)  + \cdots .
 \end{eqnarray}
 The numerical data in Fig.~\ref{fig:cond_QED} reveals that the longitudinal conductivity deviates from the chiral limit a lot more than the transverse conductivity.  This implies that the Taylor coefficients $ f_{\perp,i}$ ($i=1,2,\ldots$) are small compared to their longitudinal counterparts $ f_{\parallel,i}$ ($i=1,2,\ldots$). 
 
 As shown in Fig.~\ref{fig:cond_QED}, the strong anisotropy between the transverse and longitudinal conductivities diminishes with a decreasing magnetic field, but it does not completely vanish, even for the smallest values of $|eB|/T^2$ displayed. While the two conductivities are expected to converge in the zero magnetic field limit, this limit is nontrivial. In particular, when only one-to-two and two-to-one processes are considered and two-to-two processes are neglected, the damping rates $\Gamma_n$ must approach zero  as $B \to 0$. As a result, both conductivities formally diverge, though not necessarily at the same rate. The anisotropy will only completely disappear in the true zero-field limit, where nonzero rates $\Gamma_n$ arise from two-to-two processes.

 From a technical viewpoint, it is crucial to note that the partial Landau-level contributions to the longitudinal and transverse conductivities in Eqs.~(\ref{conductivity-perp}) and (\ref{conductivity-parallel}) behave differently. While each term in $\sigma_{\parallel}$ is proportional to $1/\Gamma_n$ and diverges as $B \to 0$, the analogous contributions to $\sigma_{\perp}$ are proportional to $\Gamma_n$, which formally vanish. However, the sum over infinitely many Landau levels in $\sigma_{\perp}$ still diverges. Thus, both $\sigma_{\parallel}$ and $\sigma_{\perp}$ diverge, albeit at different rates, accounting for the relatively high anisotropy that persists even at $|eB|/T^2\simeq 0.1$. As we will demonstrate in the following subsection, accurately calculating the conductivities for smaller values of $|eB|/T^2$ requires summing over many more Landau levels than currently used.

\subsection{Convergence of Landau-level contributions}
 
 We find that the number of Landau levels significantly contributing to $\tilde{\sigma}_{\perp}$ and $\tilde{\sigma}_{\parallel}$ increases as $|eB|/T^2$ decreases. This behavior is expected because the energy separation between Landau levels decreases with a weaker magnetic field, and the number of occupied levels grows with rising temperature. Both effects imply that more Landau levels contribute to transport. However, it is instructive to investigate in greater detail the convergence of the Landau-levels sums in both transverse and longitudinal conductivities, as defined by Eqs.~(\ref{conductivity-perp}) and (\ref{conductivity-parallel}). 
  
\begin{figure*}[hbt]
  \centering
  \includegraphics[width=0.98\columnwidth]{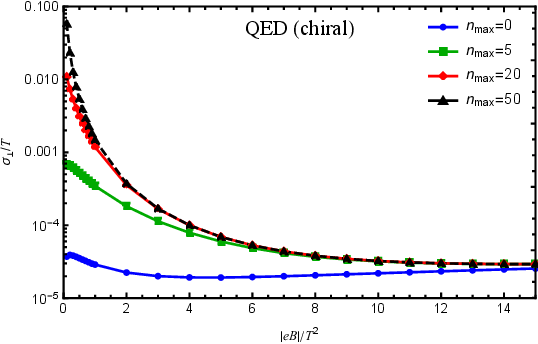} 
  \hspace{0.05\columnwidth}
  \includegraphics[width=0.98\columnwidth]{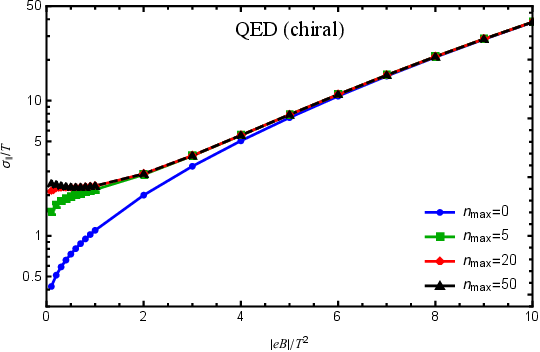} 
  \caption{Comparison of the numerical results for the transverse (left) and longitudinal (right) conductivities when the sum over Landau levels is truncated at different values of $n_{\rm max}$.}
  \label{fig:cond_QED_nmax}
 \end{figure*}

The corresponding numerical results are summarized in Fig.~\ref{fig:cond_QED_nmax}, where we present the conductivities calculated by using different truncations of the Landau-levels sums: $n_{\rm max}=0$ (blue), $n_{\rm max}=5$ (green), $n_{\rm max}=20$ (red), and $n_{\rm max}=50$ (black). Note that the rates $\Gamma_{f,n}(k_z)$ themselves are calculated by accounting for all relevant quantum transitions and full kinematic details. For simplicity, here we use the QED plasma in the chiral limit. The convergence in the case of massive QED should be comparable, if not better. 
 
 Fig.~\ref{fig:cond_QED_nmax} illustrates that the longitudinal conductivity is dominated by the lowest Landau level contribution when $b\gtrsim 5$. For the transverse conductivity, however, the analogous approximation with $n_{\rm max}=0$ works well only when $b\gtrsim 15$. It is crucial to emphasize that, in our analysis, setting $n_{\rm max}=0$ in the expression for $\sigma_\perp$ is not equivalent to the lowest Landau level approximation. Instead, it describes charge transport mediated by quantum transitions between the zeroth and first Landau levels. Such a fundamentally different mechanism is also likely responsible for the reduced sensitivity of $\sigma_\perp$ to the fermion mass as the chiral limit is approached.

 Generally, for small values of $b$, a large number of Landau levels contribute to both components of the conductivity. Based on the numerical data in Fig.~\ref{fig:cond_QED_nmax}, we estimate that one should use $n_{\rm max}\gtrsim 30/b$ and $n_{\rm max}\gtrsim 10/b$ to reduce the error to a few percent in the calculation of the transverse and longitudinal conductivities, respectively. This suggests that the approximation with $n_{\rm max}=50$ Landau levels should be reliable only down to about $b\approx 0.6$ for $\sigma_{\perp}$ and about $b\approx 0.2$ for $\sigma_{\parallel}$. As $b$ decreases beyond those values, the accuracy of the approximation gradually deteriorates. 
 
 In principle, as we argued in the previous section, the validity of the leading-order approximation for the fermion damping rate can be extended down to $b \gtrsim 0.01$. Therefore, the same should hold true for the conductivity. However, to achieve sufficiently high precision for $\sigma_{\perp}$ and $\sigma_{\parallel}$ at values of $b$ as low as $0.01$, it may be necessary to include hundreds, if not thousands, of Landau levels. While computationally more challenging, such calculations can be performed in principle.

 \subsection{Temperature and magnetic field dependence of conductivity}

\begin{figure*}[hbt]
  \centering
  \includegraphics[width=0.98\columnwidth]{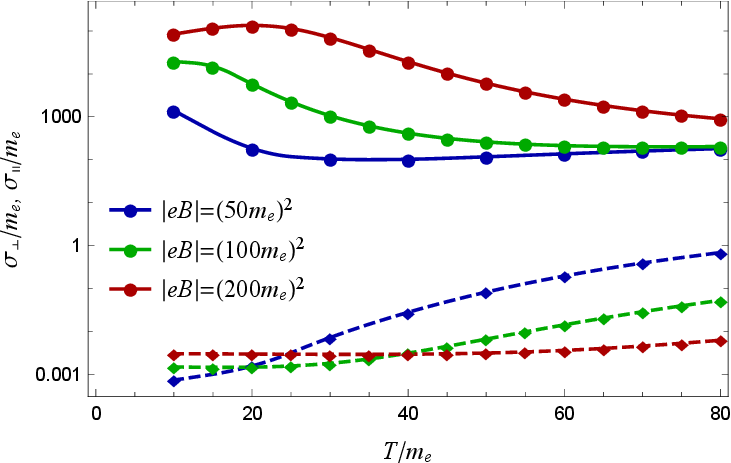} 
  \hspace{0.05\columnwidth}
  \includegraphics[width=0.98\columnwidth]{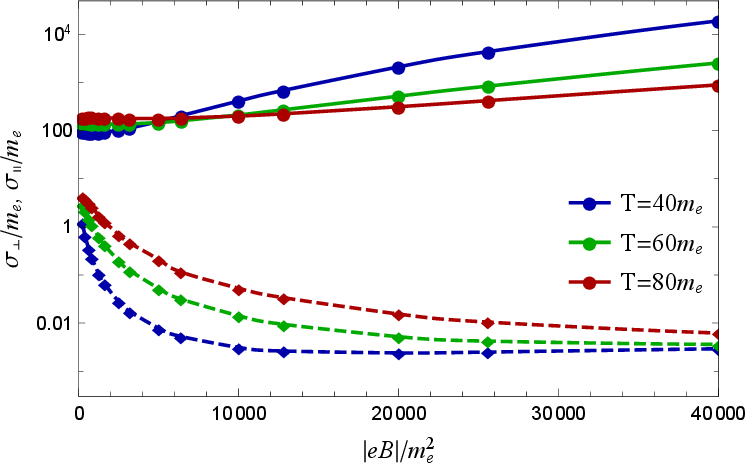} 
  \caption{The transverse (dashed lines) and longitudinal (solid lines) conductivities as functions of temperature (left) and magnetic field (right) for three fixed choices of the magnetic field (left) and three fixed temperatures (right), respectively.}
  \label{fig:cond_QED_vs-TB}
 \end{figure*}

 While the dependence of $\sigma_{\perp}/T$ and $\sigma_{\parallel}/T$ on the dimensionless ratio  $|eB|/T^2$ in Fig.~\ref{fig:cond_QED} is interesting from a theoretical viewpoint, it might be even more instructive to explore how temperature and magnetic field affect the absolute values of the transverse and longitudinal conductivities. Several representative sets of such dependences are shown in Fig.~\ref{fig:cond_QED_vs-TB}. The temperature dependence of $\sigma_{\perp}$ (dashed lines) and $\sigma_{\parallel}$ (solid lines) in the left panel reveals several interesting features. For a fixed magnetic field and sufficiently high temperatures, $T\gtrsim 0.2\sqrt{|eB|}$, the transverse conductivity tends to increase with temperature, resembling the behavior observed in conventional semiconductors. Conversely, the longitudinal conductivity decreases with temperature, similar to the behavior seen in metals.
 
 The analogy between the temperature dependence of $\sigma_{\perp}$ and $\sigma_{\parallel}$ in a magnetized plasma and the well-known qualitative behavior of the conductivity in semiconductors and metals, respectively, is useful to consider. The effect of a magnetic field, which inhibits transverse transport in the plasma, superficially resembles the effect of a gap between the valence and conduction bands in semiconductors. In both cases, raising the temperature increases electrical conductivity. In semiconductors, this increase is due to a higher density of charge carriers in the conduction band induced thermal excitations. In a magnetized plasma, raising the temperature tends to enhance the transition rates between Landau orbits, which are responsible for transverse transport. On the other hand, the longitudinal conductivity in a magnetized plasma is suppressed by thermal effects because they increase particle scattering. This is essentially the same mechanism that reduces conductivity in metals as temperature rises. 
 
 The behavior of $\sigma_{\perp}$, shown in the left panel of Fig.~\ref{fig:cond_QED_vs-TB}, may seem puzzling at low temperatures. In a finite range of parameters, namely $T\lesssim 0.2\sqrt{|eB|}$, the transverse conductivity remains nearly independent of the temperature before starting to increase at higher values. Additionally, the ordering of conductivities at fixed magnetic fields reverses compared to their ordering at high $T$. The low-temperature regime roughly corresponds to $b\gtrsim 2b_{{\rm min},\perp}\approx 25$ in Fig.~\ref{fig:cond_QED}. In this range, transitions between the lowest and first Landau levels dominate the dimensionless conductivity $\tilde{\sigma}_{\perp}$. As discussed earlier, it increases as the square root of $b$, i.e., $\tilde{\sigma}_{\perp}\simeq 1.5\times 10^{-5} \sqrt{b}$. Therefore, the corresponding dimensionful conductivity remains nearly constant, i.e., $\sigma_{\perp}\equiv T \tilde{\sigma}_{\perp} \simeq 1.5\times 10^{-5}\sqrt{|eB|}$ when $T\lesssim 0.2\sqrt{|eB|}$.  
 
 The dependence of the conductivities $\sigma_{\perp}$ and $\sigma_{\parallel}$  on the magnetic field is presented in the right panel in Fig.~\ref{fig:cond_QED_vs-TB}. The transverse and longitudinal conductivities are interpolated by dashed and solid lines, respectively. The three data sets correspond to different values of the temperature, $T=40m_e$ (blue), $T=60m_e$ (green), and $T=80m_e$ (red). In accordance with the general expectations, $\sigma_{\perp}$ has the tendency to go down and $\sigma_{\parallel}$  to go up with increasing the magnetic field. The transverse conductivity is suppressed by the magnetic field because it enhances trapping of the charged carriers in Landau-level orbits. On the other hand, the longitudinal conductivity becomes higher at large $|eB|$ because the damping rates get suppressed and the density of states in the Landau levels grows linearly with the field.  
 
 Although our primary focus was on the ultrarelativistic regime, qualitatively similar temperature and magnetic field dependences of the transverse and longitudinal conductivities can be observed in regimes with much weaker fields ($m_e \lesssim \sqrt{|eB|} \lesssim 5m_e$) and much lower temperatures ($m_e \lesssim T \lesssim 5m_e$). Interestingly, while the absolute values of the two transport coefficients differ significantly, the dimensionless characteristics  $\tilde{\sigma}_{\perp}$ and $\tilde{\sigma}_{\parallel}$ closely follow the scaling behavior shown in Fig.~\ref{fig:cond_QED}, provided $b\lesssim 1$. At larger $b$ values, both $\tilde{\sigma}_{\perp}$ and $\tilde{\sigma}_{\parallel}$ tend to be smaller than their scaling values in Fig.~\ref{fig:cond_QED}. Specifically, $\tilde{\sigma}_{\perp}$ becomes quantitatively much smaller with increasing $b$, whereas deviations for $\tilde{\sigma}_{\perp}$ remain moderate ($\lesssim 25\%$) even for $b\simeq 25$.

 \subsection{Electrical conductivity at nonzero chemical potential}
 
 It is instructive to discuss briefly the effect of a nonzero chemical potential $\mu$ on the electrical conductivity. In the case of a QED plasma, including a nonzero $\mu$ has significant implications. Most importantly, having a nonzero chemical potential for the electrons implies that the plasma is not neutral overall unless there are also protons (or other positively charged particles) to compensate for the negative charge of the electrons. Here, we will ignore this complication and analyze only the partial contribution of electrons to the conductivity. This simplification is justified at leading order in coupling when only one-to-two and two-to-one processes contribute to the damping rate. Indeed, electron scattering on protons is a two-to-two process that contributes to the damping rate at the subleading order in coupling.
 
 A nonzero chemical potential $\mu$ affects conductivity in two main ways. Firstly, it modifies particle distributions in the plasma, which appear explicitly in the definition of the conductivities in Eqs.~(\ref{conductivity-perp}) and (\ref{conductivity-parallel}). Secondly, it changes the  fermion damping rates. In this analysis, we account for the former but not the latter. In other words, we will ignore the direct dependence of the damping rates on the chemical potential, assuming that it is less significant at small values of $\mu$. 
 
\begin{figure*}[t]
	\centering
	\includegraphics[width=0.98\columnwidth]{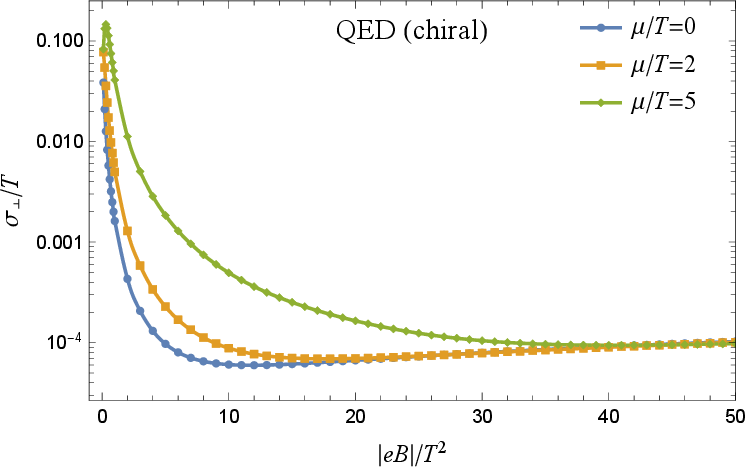} 
	\hspace{0.05\columnwidth}
	\includegraphics[width=0.98\columnwidth]{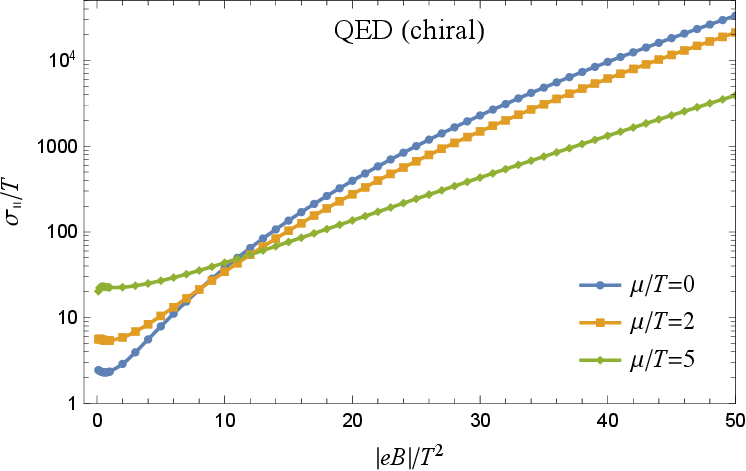} 
	\caption{The transverse (left) and longitudinal (right) conductivities in units of temperature as functions of the dimensionless ratio $|eB|/T^2$ for three different choices of the electron chemical potential: $\mu/T=0$ (blue),  $\mu/T=2$ (orange), and $\mu/T=5$ (green).}
	\label{fig:cond_QED_mu}
\end{figure*}

 Furthermore, for simplicity, we will consider only the case of chiral QED, where the chemical potential introduces one additional dimensionless parameter, $\mu/T$. The corresponding numerical data for the transverse and longitudinal conductivities as functions of the dimensionless ratio $|eB|/T^2$ are presented in Fig.~\ref{fig:cond_QED_mu}. We show the results for three different values of the electron chemical potential: $\mu/T=0$ (blue),  $\mu/T=2$ (orange), and $\mu/T=5$ (green). 
 
 The data in Fig.~\ref{fig:cond_QED_mu} reveals that the electrical chemical potential has very different effects on the transverse and longitudinal conductivities. The transverse conductivity generally increases with the addition of a nonzero chemical potential. This effect is localized primarily in a finite region at small values of $|eB|/T^2$, but the width of this region grows as $\mu$ increases. The impact on the longitudinal conductivity is different. It is enhanced by the chemical potential  when $|eB|/T^2$ is small but suppressed when $|eB|/T^2$ is large. 
 
 When $|eB|/T^2 \lesssim 5$, the effect of $\mu$ on the longitudinal conductivity is similar to that in metals. It expands the phase space of charge carriers that fill many Landau levels. Roughly speaking, this is analogous to an increasing density of states at the Fermi surface when $B=0$, resulting in an enhanced value of $\tilde{\sigma}_{\parallel}$. In contrast, $\tilde{\sigma}_{\parallel}$ tends to be suppressed by $\mu$ at large $|eB|/T^2$. In this regime, only the lowest Landau level contributes to the conductivity, and its dimensional reduction (from $3$ to $1$ spatial directions) makes the effect of a moderate $\mu$ negligible. As in one-dimensional systems, the density of states at the Fermi surface is independent of $\mu$. The only effect of $\mu$ is to shifts the phase space from being centered $k_z\approx 0$ to $k_z\approx \mu$. Since the damping rate of charge carriers tends to increase with $k_z$, their contribution to the conductivity gets suppressed. 
 
 A different mechanism behind transverse transport explains why $\mu$ has a negligible effect on $\tilde{\sigma}_{\perp}$ at large $|eB|/T^2$. Indeed, transverse conductivity is driven primarily by quantum transitions between the lowest and first Landau levels. Recall that we neglected the direct effect of a nonzero chemical potential on the damping rates. Regarding the effect on particle distributions, they affect $\tilde{\sigma}_{\perp}$ only minimally when $|eB|/T^2\gg 1$.

 \section{Electrical conductivity in QGP}
 \label{QGPresults}
 
 Let us now turn to the case of strongly magnetized QGP made of the lightest up and down quarks. Unlike weakly coupled QED plasma, QGP consists of strongly interacting quarks. Due to this strong coupling, most perturbative techniques cannot be rigorously applied to study such a system. However, at sufficiently high temperatures, QGP is expected to become weakly interacting due to the QCD property of asymptotic freedom. Therefore, in this study, we assume the temperature is high enough to justify using the leading-order approximation in calculating the quark damping rates.
 
 Without loss of generality, we assume the masses of both up and down quarks are the same, $m =5~\mbox{MeV}$. However, we account for their different electrical charges, $q_u=2e/3$ and $q_d=-e/3$, which affect their interaction with the background magnetic field. 
 
 As in the QED study in the previous section, we focus primarily on the regime of high temperatures ($T\gg m$) and strong magnetic fields ($|eB|\gg m^2$), where the effects of nonzero quark masses are small. This regime is particularly relevant for heavy-ion collisions, where deconfined QGP is produced. Typical temperatures of such plasma are well above the deconfinement value, $T_c\simeq 160~\mbox{MeV}$, and the magnetic field is of the order of $|eB|\gtrsim m_{\pi}^2$ according to many theoretical estimates  
 \cite{Skokov:2009qp,Voronyuk:2011jd,Deng:2012pc,Bloczynski:2012en,Guo:2019mgh}. It is worth noting, however, that the rapid evolution of QGP fireballs can drastically reduce the background magnetic field in the plasma  \cite{McLerran:2013hla,Tuchin:2015oka,Yan:2021zjc}. 
 
 Similarly to the QED case, the quark damping rates in QGP are determined by one-to-two and two-to-one processes. However, in the corresponding processes, photons should be replaced by gluons. Therefore, when calculating $\Gamma_{f,n}^{\rm QGP}(k_z)$ using Eq.~(\ref{Gamma_n_pz-short}), we should replace the coupling constant $\alpha_{*}$ with $\alpha_s C_F$, where $\alpha_s$ is the strong coupling constant, $C_F=(N_c^2-1)/(2N_c)=4/3$, and $N_c=3$. Note that the strong coupling constant at temperatures relevant for heavy-ion physics is expected to be of the order of $1$.  
 
 Considering the ultrarelativistic regime of QGP, it is convenient to express conductivities in units of temperature, $\tilde{\sigma}_{\perp}\equiv \sigma_{\perp}/T$ and $\tilde{\sigma}_{\parallel}\equiv \sigma_{\parallel}/T$. It is reasonable to expect that the dimensionless conductivities $\tilde{\sigma}_{\perp}$ and $\tilde{\sigma}_{\parallel}$, can be approximated by some scaling functions of the ratio $|eB|/T^2$. 
 
\begin{figure}[t]
  \centering
  \includegraphics[width=\columnwidth]{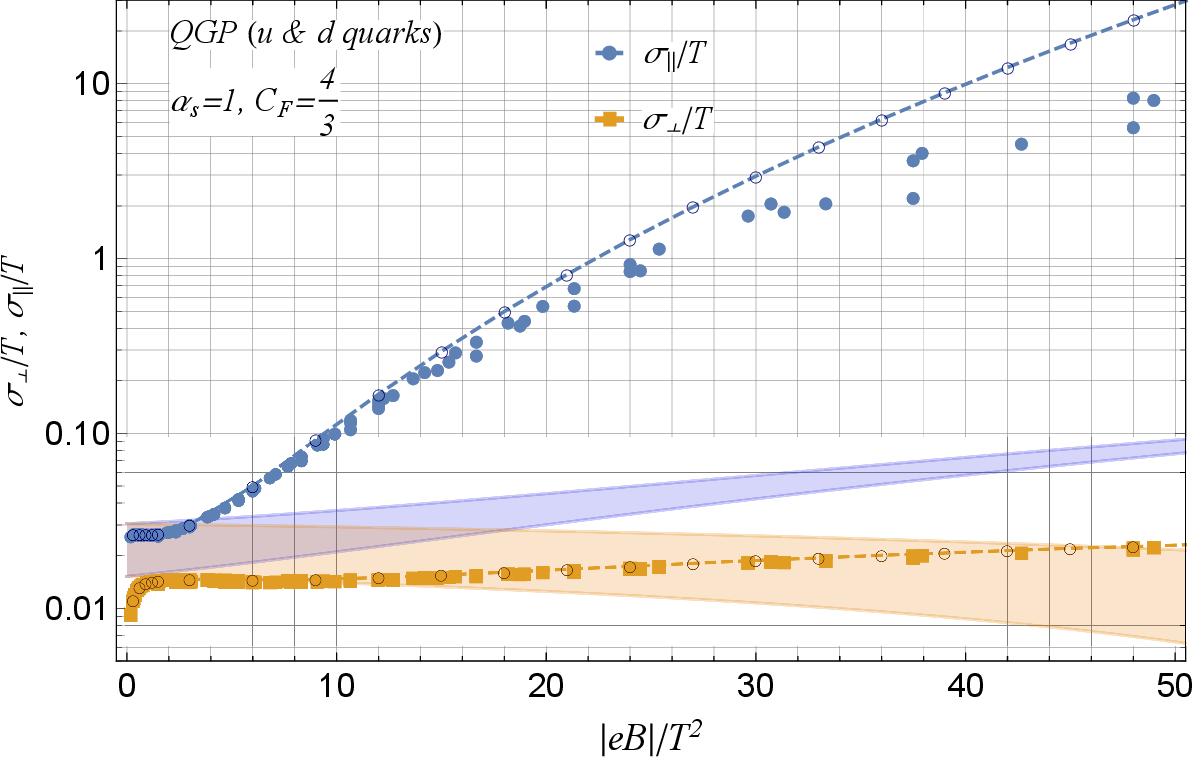} 
  \caption{The transverse and longitudinal conductivities in two-flavor QGP in units of temperature as functions of the dimensionless ratio $|eB|/T^2$.  Empty circles and interpolating dashed lines represent the conductivities in the chiral limit.}
  \label{fig:cond_QCD}
 \end{figure}

 Our numerical data for the QGP conductivities are presented in Fig.~\ref{fig:cond_QCD}, which includes a large set of data points for a wide range of temperatures, $0.85 m_\pi \lesssim T \lesssim 3 m_{\pi}$ and magnetic fields, $0.3 m_\pi^2 \lesssim |eB| \lesssim 82 m_\pi^2 $. The  numerical values of the conductivities are also provided as data files in the Supplemental Material \cite{QED-QCDCond:2024}. For these calculations, we used $\alpha_s=1$. While such a large value may be outside the validity range of the leading-order approximation, it can still provide useful qualitative estimates for the conductivities in QGP.
 
 Following the same approach as in the QED case in the previous section, we combine the results for both massive and massless (chiral) versions of QGP  in Fig.~\ref{fig:cond_QCD}. Filled markers correspond to the QGP with a nonzero (physical) quark mass, while empty circles represent data in the chiral limit of QGP. To provide a visual guide, the chiral data are interpolated by dashed lines. 
 
 We find that the transverse and longitudinal conductivities in the chiral limit of QGP can be well approximated by the following Pad\'{e} approximants:
  \begin{widetext}
  \begin{eqnarray}
  \tilde{\sigma}_{\perp, {\rm chiral}}^{\rm QCD} &\simeq& \tilde{\sigma}_{\perp,0}
  \frac{ 1 + 12.5 b + 0.643 b^2 + 0.08 b^3 + 0.00081 b^4 - 9.8 \times 10^{-7} b^5}{1 + 3.7 b + 0.35 b^2 + 0.021 b^3}, 
  \label{signa11QCDfit}  \\
  \tilde{\sigma}_{\parallel, {\rm chiral}}^{\rm QCD}  &\simeq& \tilde{\sigma}_{\parallel,0}
  \frac{ 1 - 0.06 b + 0.0292 b^2 - 10^{-4} b^3 + 6.2 \times 10^{-5} b^4 + 1.32 \times 10^{-8} b^6}{1 - 0.0087 b +  2.4 \times 10^{-5} b^2 - 1.67 \times 10^{-8} b^3} ,
  \label{signa33QCDfit}
 \end{eqnarray}
  \end{widetext}
  where $\tilde{\sigma}_{\perp,0}\approx 5\times 10^{-3}$, $\tilde{\sigma}_{\parallel,0} \approx 2.7 \times 10^{-2}$, and $b=|eB|/T^2$. The quality of the fit is better than about $2\%$ for the transverse and $10\%$ for the longitudinal conductivity in the whole range of parameters studied, $|eB|/T^2\lesssim 60$.
 
 For comparison, in Fig.~\ref{fig:cond_QCD}, we also include the QCD lattice results for the electrical conductivities obtained in Ref.~\cite{Astrakhantsev:2019zkr}. They are represented by the orange and blue shaded bands, where the edges of the bands correspond to different choices of the zero-field conductivities, i.e., $\sigma_0=0.3 C_{\rm em}T$ and $\sigma_0=0.6 C_{\rm em} T$, respectively, with $C_{\rm em} = 5e^2/9$.
 
 As seen, our numerical data for the transverse conductivity are of the same order of magnitude as the lattice results, although the general profile of its dependence on $|eB|/T^2$ looks different. In contrast, the longitudinal conductivity is much larger compared to the lattice data. We can only speculate that this difference is due to higher-order processes contributing significantly to the damping rates and, more generally, the spectral densities of quarks. The longitudinal conductivity  appears to be quite sensitive to the damping rates even in the case of QED. It is reasonable to expect that this sensitivity is exacerbated in a strongly coupled QGP.
 
 Since the value of the QCD coupling constant $\alpha_{s}$  is not well known under conditions relevant for heavy-ion physics, it is useful to examine how varying $\alpha_{s}$ affects the transverse and longitudinal conductivities. The corresponding numerical data for two different values, $\alpha_{s}=0.5$ and  $\alpha_{s}=2$, are shown in Fig.~\ref{fig:cond_QCD-alphas}. The values of the conductivities can also be found in the Supplemental Material \cite{QED-QCDCond:2024}. 
 
  \begin{figure*}[htb]
  \centering
  \includegraphics[width=0.98\columnwidth]{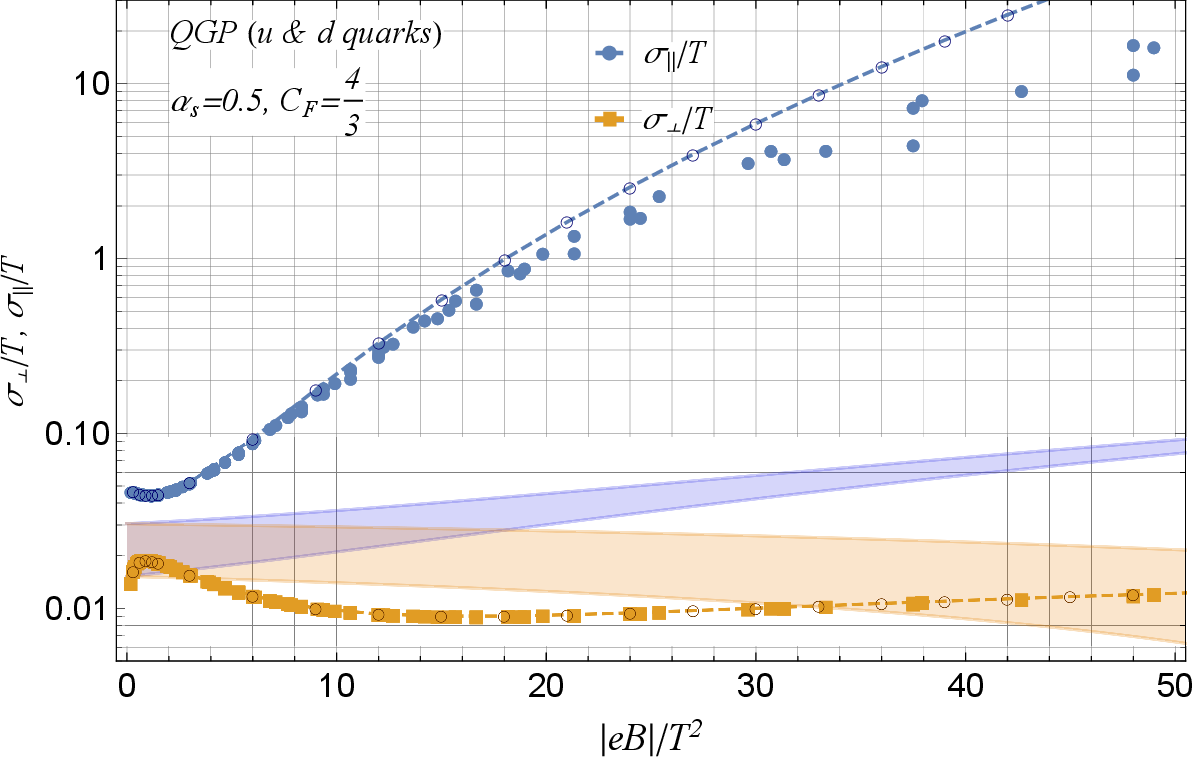}
  \hspace{0.05\columnwidth}
  \includegraphics[width=0.98\columnwidth]{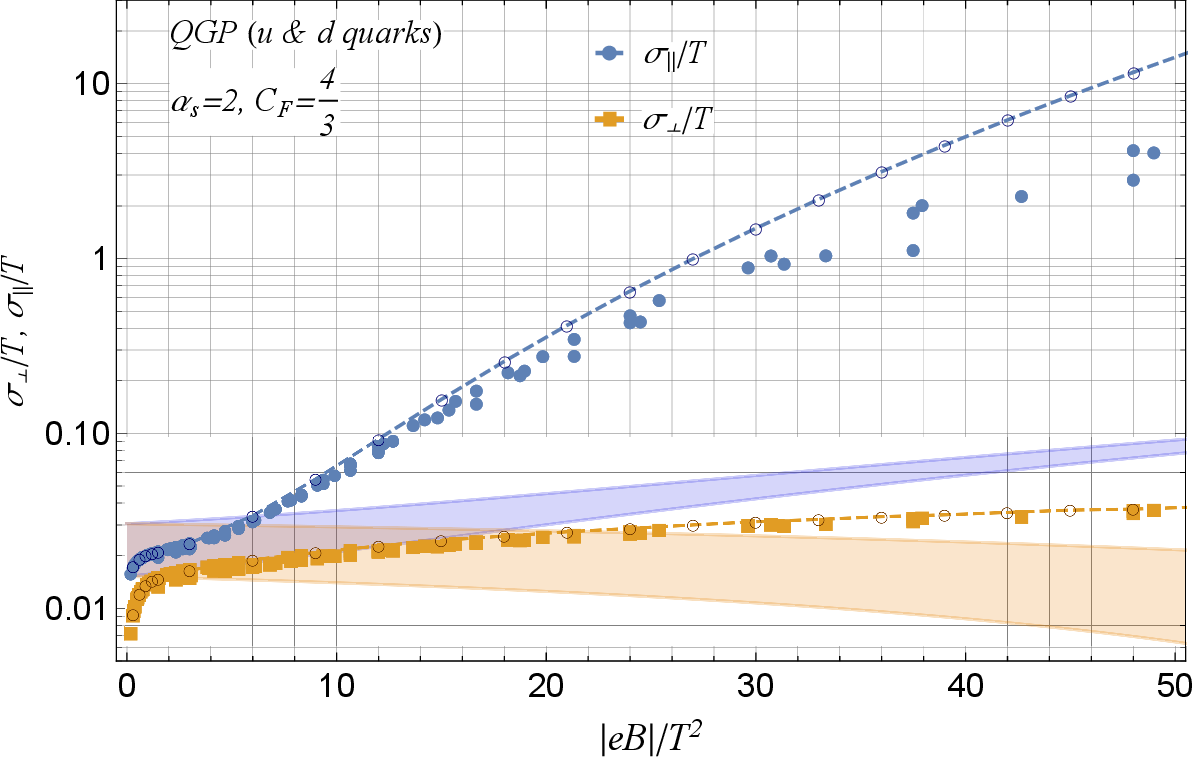} 
  \caption{The transverse and longitudinal conductivities in two-flavor QGP for two different choices of strong coupling:  $\alpha_{s}=0.5$ (left)
   and $\alpha_{s}=2$ (right). Empty circles and interpolating dashed lines represent the data in the chiral limit. The blue and orange bands show the approximate ranges of  transverse and longitudinal conductivities obtained in lattice QCD \cite{Astrakhantsev:2019zkr}.}
  \label{fig:cond_QCD-alphas}
 \end{figure*}
 
 By comparing the data in the two panels of Fig.~\ref{fig:cond_QCD-alphas}, we see that the two conductivities get closer together as the coupling constant increases. This is consistent with the underlying mechanisms of the transverse and longitudinal conductivities: while the former is proportional to the damping rate, the latter is inversely proportional to it, and thus to the strong coupling constant. This suggests that the ratio $\sigma_{\perp}/\sigma_{\parallel}$, which quantifies the anisotropy of electrical charge transport, should scale as $\alpha_s^2$. Of course, the real situation is more complex because the conductivities are affected very differently by temperature and magnetic field.

 \section{Discussion and Summary}
 \label{Summary}
 
 In this paper, we studied the transverse and longitudinal electrical conductivities of strongly magnetized relativistic plasmas within the gauge field theory framework. We relied on the Kubo formula and detailed knowledge of the fermion damping rates of charge carriers in the Landau-level representation. In the regime of strong fields, the leading-order results for the damping rates are determined by the following three types of one-to-two and two-to-one processes: $\psi_{n}\to \psi_{n^\prime}+\gamma$, $\psi_{n}+\gamma \to \psi_{n^\prime}$,  and $\psi_{n}+\bar{\psi}_{n^\prime}\to \gamma$. The corresponding rates were calculated using full kinematics and exact amplitudes in Ref.~\cite{Ghosh:2024hbf}. We argue that such an approximation is reliable when the magnetic field and temperature satisfy the inequality $|eB|/T^2\gg \alpha $. In this regime, the subleading two-to-two processes are suppressed. It is important to keep in mind, however, that they would become dominant when the magnetic field becomes sufficiently weak, i.e., $|eB|\lesssim \alpha T^2$. In such a weak-field regime, the importance of the Landau-level quantization diminishes. 
 
 We investigated in detail the temperature and magnetic field dependence of the electric charge transport. In the ultrarelativistic regime, the transverse and longitudinal conductivities are approximated by universal scaling functions of a single dimensionless parameter $|eB|/T^2$. The shape of the corresponding functions is established by considering the chiral limit of QED plasma, which assumed that the fermion mass vanishes. When the mass is nonzero, the scaling appears to work really well only for the transverse component of conductivity. The longitudinal conductivity, on the other hand, shows significant deviations from the chiral limit even for relatively large values of $T/m_e$, especially when $|eB|/T^2 \gtrsim 10$. 
 
 For a QED plasma, the corresponding numerical results are presented in Fig.~\ref{fig:cond_QED}. We find that underlying mechanisms of charge transport are very different in the transverse and longitudinal directions with respect to the magnetic field. The nature of longitudinal transport is similar to that in metals, where an increasing damping (or scattering) rate, e.g., due to raising the plasma temperature, suppresses the conductivity. Transverse transport, on the other hand, is somewhat similar to that in semiconductors, where the conductivity grows with temperature. The reason for such a behavior of the transverse conductivity is an entrapment of charge carriers in Landau orbits. In essence, this is the same argument of a dimensional reduction, $3+1 \to 1+1$, that one uses in the theory of magnetic catalysis \cite{Miransky:2015ava}. A strong magnetic field restricts the transverse motion of charged fermions down to spatial scales of the order of the magnetic length $\ell=1/\sqrt{|eB|}$. As a result, the transverse conductivity happens only as a result of transitions between Landau levels. Therefore, the conductivity grows with rising temperature because it enhances the corresponding transition rates.
 
 The drastically different mechanisms of the transverse and longitudinal transport imply that conductivity is highly anisotropic in a strong magnetic field. This effect is amplified when the coupling constant $\alpha$ is small. The transverse conductivity is proportional to the damping rate, whereas the longitudinal conductivity is inversely proportional to the damping rate, resulting in the ratio $\sigma_{\perp}/\sigma_{\parallel}$ scaling as $\alpha^{2}$. In the case of QED, we find that the values of $\sigma_{\perp}$ are typically $2$ to $7$ orders of magnitude smaller than $\sigma_{\parallel}$. 
 
It should be noted that our study goes well beyond the previous attempts to calculate $\sigma_{\parallel}$ in Refs.~\cite{Hattori:2016cnt,Hattori:2016lqx}, which relied on a naive lowest Landau-level approximation. Formally, our analytical result appears to agree with those in Refs.~\cite{Hattori:2016cnt,Hattori:2016lqx}, provided $\sigma_{\parallel}$ is implicitly expressed in terms of the damping rate $\Gamma_0(k_z)$. However, as we discussed in Sec.~\ref{subsec:Scaling-chiral}, the inclusion of the transitions only within the lowest Landau level is insufficient for the calculation of the damping rate $\Gamma_{f,0}(k_z)$, even under very strong magnetic fields. This limitation is particularly important when the fermion mass is small and the longitudinal momentum is large.
 
In this study, we utilized recent state-of-the-art results for the fermion self-energy in the Landau-level representation \cite{Ghosh:2024hbf} to calculated the damping rates, properly accounting for contributions from all Landau levels. This approach was crucial for obtaining reliable results for the longitudinal conductivity at leading order in coupling, as well as for calculating the transverse conductivity $\sigma_{\perp}$ for the first time within a gauge theory. In comparison, the transverse conductivity is not even accessible in the kinetic theory approach used in Refs.~\cite{Hattori:2016lqx,Fukushima:2017lvb,Fukushima:2019ugr}.
 
It is worth noting that the magnetic field dependence of the longitudinal conductivity is consistent with the qualitative results for negative magnetoresistance observed in Dirac and Weyl semimetals \cite{Kim:2013ys,Li:2014bha,Xiong:2015yq,Feng:2015PRB92,Li-Yu-Cd3As2:2015,Li-Wang-Cd3As2:2016,Huang:2015ve,Zhang:2015gwa}. This reconfirms the same conclusion reached in Ref.~\cite{Fukushima:2017lvb,Fukushima:2019ugr} that a self-consistent treatment of electrical charge transport within gauge theory is sufficient to reproduce negative magnetoresistance, without the need for additional phenomenological assumptions about the interplay between the chiral magnetic effect and chiral charge diffusion. This should not be surprising, of course, since the quantum chiral anomaly is inherently embedded in our first principles field-theoretic analysis. Furthermore, this is consistent with the claim made in Ref.~\cite{Ambjorn:1983hp} that the anomaly is entirely captured by the topological nature of the lowest Landau level. Beyond reproducing longitudinal negative magnetoresistance, our analysis of the transverse conductivity also enables us to calculate magnetoresistivity for any arbitrary angle between the magnetic field and the electric current.
 
 We additionally investigated the effects of a nonzero chemical potential $\mu$ on the transverse and longitudinal  conductivities. We found that both transport characteristics tend to increase with the chemical potential when $|eB|/T^2\lesssim 5$. This effect is similar to that observed in metals in condensed matter physics, where conductivity increases due to a higher density of states at the Fermi surface. However, at large $|eB|/T^2$, the longitudinal conductivity is suppressed by $\mu$ due to an increasing damping rate for the relevant states at a dimensionally reduced Fermi surface. In contrast, the transverse conductivity, which is determined by the rate of transitions between the zeroth and first Landau levels, remains almost insensitive to $\mu$ at large $|eB|/T^2$.
 
 It should be noted that when studying a magnetized plasma at a nonzero chemical potential, we accounted only for the effects of $\mu$ on particle distributions, not on fermion damping rates. The latter are expected to become increasingly important as the chemical potential increases. Before conducting a corresponding study, it is necessary to generalize the results for fermion damping rates in Ref.~\cite{Ghosh:2024hbf} to nonzero $\mu$. This is a problem we plan to address in the future.
 
We have extended our study of electrical conductivity from a weakly coupled QED plasma to a strongly coupled QGP in the presence of a background magnetic field. Our key findings are presented in Figs.~\ref{fig:cond_QCD} and \ref{fig:cond_QCD-alphas}, where we used three different values of the coupling constant $\alpha_s$. Although the leading-order results might not be as reliable as those in QED, we believe that our model accurately captures the qualitative physics of charge transport. We hope our findings will help guide research in heavy-ion physics and identify qualitative features of QGP associated with a strong magnetic field. With that said, further theoretical research is needed to investigate the role of subleading processes and nonperturbative effects in the charge transport. 
 
 Electrical conductivity is crucial for QGP produced in heavy-ion collisions, as it determines the timescales for the trapping and diffusion of the magnetic field and may influence observed charged particle correlations. Recently, the STAR collaboration provided the first experimental results for conductivity, as reported in Ref.~\cite{STAR:2023jdd}. We hope that future experiments will detect the effects of a nonzero magnetic field in the measured transport characteristics of QGP. 
 
 Reliable results for the conductivity of the electron-positron plasma are crucial for realistic simulations of pulsar magnetospheres \cite{Li:2011zh,Kalapotharakos:2011vg,Kalapotharakos:2012dq}. Specifically, the inverse of $\sigma_{\parallel}$ quantifies the deviation from the force-free condition in the plasma, the magnitude of the parallel component of the electric field, and the electrical current dissipation rate. These factors determine the energy balance and activity of the magnetosphere, which, in turn, affect the observational signatures of pulsars. Therefore, our results are likely to be useful for a better understanding of magnetosphere physics and interpreting pulsar activity.
 
\acknowledgments{This research was funded in part by the U.S. National Science Foundation under Grant No.~PHY-2209470.}
 
 \appendix
 
 \section{Dirac traces}
 \label{app:traces}
 
 To calculate individual components of the conductivity tensor in Eq.~(\ref{sigma-tensor-spectral}), we used the following results for Dirac traces:
  \begin{widetext}
  \begin{eqnarray}
  T_{aa}^{11} &=& \mbox{tr} \big[ 
  \gamma^1 \left(E_{n} \gamma^{0} -\lambda  k_{z}\gamma^3+\lambda  m_{f} \right)
  \left({\cal P}_{+}L_n -{\cal P}_{-}L_{n-1} \right) \gamma^1 \left(E_{n^\prime} \gamma^{0} -\lambda^{\prime}  k_{z}\gamma^3+\lambda^{\prime}  m_{f} \right)
  \left({\cal P}_{+}L_{n^\prime} -{\cal P}_{-}L_{n^\prime-1} \right)  \big]
  \nonumber\\
  &=& - 2\left( L_n  L_{n^\prime-1}+L_{n^\prime}  L_{n-1}\right) \left[E_n E_{n^\prime}-\lambda \lambda^{\prime} (k_z^2+
  m_{f}^2)\right]  , \\
  T_{aa}^{12} &=&  - T_{aa}^{21} = \mbox{tr} \big[ 
  \gamma^1 \left(E_{n} \gamma^{0} -\lambda  k_{z}\gamma^3+\lambda  m_{f} \right)
  \left({\cal P}_{+}L_n -{\cal P}_{-}L_{n-1} \right)  \gamma^2 \left(E_{n^\prime} \gamma^{0} -\lambda^{\prime}  k_{z}\gamma^3+\lambda^{\prime}  m_{f} \right)
  \left({\cal P}_{+}L_{n^\prime} -{\cal P}_{-}L_{n^\prime-1} \right)  \big] \nonumber\\
  &=& 2 i s_\perp \left( L_n  L_{n^\prime-1}- L_{n^\prime}  L_{n-1}\right) \left[E_n E_{n^\prime}-\lambda \lambda^{\prime} 
  (k_z^2+m_{f}^2)\right] , \\
  T_{aa}^{33} &=& \mbox{tr} \big[ 
  \gamma^3 \left(E_{n} \gamma^{0} -\lambda  k_{z}\gamma^3+\lambda  m_{f} \right)
  \left({\cal P}_{+}L_n -{\cal P}_{-}L_{n-1} \right)   \gamma^3  \left(E_{n^\prime} \gamma^{0} -\lambda^{\prime}  k_{z}\gamma^3+\lambda^{\prime}  m_{f} \right)
  \left({\cal P}_{+}L_{n^\prime} -{\cal P}_{-}L_{n^\prime-1} \right) \big]
  \nonumber\\
  &=& 2 \left(L_nL_{n^\prime}+  L_{n-1}L_{n^\prime-1}\right) \left[E_nE_{n^\prime}+\lambda \lambda^{\prime} \left(k_z^2-
  m_{f}^2\right) \right], 
  \end{eqnarray}
  \begin{eqnarray}
  T_{bb}^{11} &=& -  T_{bb}^{22} = 4\lambda \lambda^{\prime} \mbox{tr} \left[ 
  \gamma^1 (\bm{k}_\perp\cdot\bm{\gamma}_\perp) L_{n-1}^1 
  \gamma^1 (\bm{k}_\perp\cdot\bm{\gamma}_\perp) L_{n^{\prime} -1}^1  \right]
= 16 \lambda \lambda^{\prime} \left(k_x^2-k_y^2\right) L_{n-1}^1 L_{n^{\prime} -1}^1 , \\
  T_{bb}^{12} &=& T_{bb}^{21} = 4\lambda \lambda^{\prime} \mbox{tr} \left[ 
  \gamma^1 (\bm{k}_\perp\cdot\bm{\gamma}_\perp) L_{n-1}^1 
  \gamma^2 (\bm{k}_\perp\cdot\bm{\gamma}_\perp) L_{n^{\prime} -1}^1  \right]
  =32 \lambda \lambda^{\prime}  k_x k_y L_{n-1}^1 L_{n^{\prime} -1}^1 , \\
  T_{bb}^{33}  &=& 4\lambda \lambda^{\prime} \mbox{tr} \left[ 
  \gamma^3 (\bm{k}_\perp\cdot\bm{\gamma}_\perp) L_{n-1}^1 
  \gamma^3 (\bm{k}_\perp\cdot\bm{\gamma}_\perp) L_{n^{\prime}-1}^1  \right]
  =-16 \lambda \lambda^{\prime} k_\perp^2  L_{n-1}^1 L_{n^{\prime} -1}^1, \\
  T_{ab}^{11} &=& 2\lambda^{\prime}  \mbox{tr} \left[ 
  \gamma^1 \left(E_{n} \gamma^{0} -\lambda  k_{z}\gamma^3+\lambda  m_{f} \right)
  \left({\cal P}_{+}L_n -{\cal P}_{-}L_{n-1} \right) 
  \gamma^1 (\bm{k}_\perp\cdot\bm{\gamma}_\perp) L_{n^{\prime} -1}^1  \right]=0, \\
  T_{ab}^{12} &=&  - T_{ab}^{21} = 2\lambda^{\prime}  \mbox{tr} \left[ 
  \gamma^1 \left(E_{n} \gamma^{0} -\lambda  k_{z}\gamma^3+\lambda  m_{f} \right)
  \left({\cal P}_{+}L_n -{\cal P}_{-}L_{n-1} \right) 
  \gamma^2 (\bm{k}_\perp\cdot\bm{\gamma}_\perp) L_{n^{\prime} -1}^1  \right] = 0,\\
  T_{ab}^{33} &=& 2\lambda^{\prime}  \mbox{tr} \left[ 
  \gamma^3 \left(E_{n} \gamma^{0} -\lambda  k_{z}\gamma^3+\lambda  m_{f} \right)
  \left({\cal P}_{+}L_n -{\cal P}_{-}L_{n-1} \right) 
  \gamma^3 (\bm{k}_\perp\cdot\bm{\gamma}_\perp) L_{n^{\prime} -1}^1  \right]=0 .
 \end{eqnarray}
  Note that similar traces of the Lorentz components with one transverse ($i=1,2$) and one longitudinal ($j=3$) index do not contribute to the conductivity tensor. Indeed, one can verify that  $T_{aa}^{13}= T_{aa}^{23}=T_{bb}^{13}= T_{bb}^{23}=0$, while $T_{ab}^{13} \sim (a_1 k_x + b_1  k_x ) k_z $ and $T_{ab}^{23} \sim (a_2 k_x + b_2  k_x ) k_z $. The latter two do not contribute to $\sigma_{ij}$ because they are odd functions of the spatial components of the momentum $\bm{k}$.
 
 In the derivation of the transverse and longitudinal conductivities in Eqs.~(\ref{conductivity-perp}) and (\ref{conductivity-parallel}), we also made use of the following sums over $\lambda$ and $\lambda^\prime$:
  \begin{eqnarray}
\sum_{\lambda=\pm}\sum_{\lambda^\prime=\pm}
  \frac{\Gamma_n}{\left(k_{0} -\lambda E_{n} \right)^2+\Gamma_n^2}
  \frac{\Gamma_{n^\prime}}{\left(k_{0} -\lambda^{\prime}  E_{n^\prime} \right)^2+\Gamma_{n^\prime}^2}  &=&\frac{4\Gamma_{n} \Gamma_{n^\prime}\left(k_{0}^2+E_{n}^2+\Gamma_{n}^2 \right) \left(k_{0}^2+E_{n^\prime}^2+\Gamma_{n^\prime}^2 \right) 
  }{\left[\left(E_{n}^2+\Gamma_{n}^2 -k_{0}^2\right)^2+4k_{0}^2 \Gamma_{n}^2 \right]
   \left[\left(E_{n^\prime}^2+\Gamma_{n^\prime}^2 -k_{0}^2\right)^2+4k_{0}^2 \Gamma_{n^\prime}^2 \right]},
  \\
\sum_{\lambda=\pm}\sum_{\lambda^\prime=\pm}
  \frac{\lambda\Gamma_n}{\left(k_{0} -\lambda E_{n} \right)^2+\Gamma_n^2}
  \frac{\lambda^{\prime} \Gamma_{n^\prime}}{\left(k_{0} -\lambda^{\prime}  E_{n^\prime} \right)^2+\Gamma_{n^\prime}^2}  &=&\frac{16k_0^2 E_{n} E_{n^\prime} \Gamma_{n} \Gamma_{n^\prime}}{\left[\left(E_{n}^2+\Gamma_{n}^2 -k_{0}^2\right)^2+4k_{0}^2 \Gamma_{n}^2 \right]
   \left[\left(E_{n^\prime}^2+\Gamma_{n^\prime}^2 -k_{0}^2\right)^2+4k_{0}^2 \Gamma_{n^\prime}^2 \right]}.
 \end{eqnarray}
   \end{widetext}

 \section{Amplitude of leading-order processes}
 \label{app:amplitude}
 
 As we stated in the main text, the damping rates in the Landau-level states, $\Gamma_{f,n}(k_z)$, are determined by three types of one-to-two and two-to-one processes: (i) $\psi_{n}\to \psi_{n^\prime} +\gamma$, (ii) $\psi_{n}+\gamma\to \psi_{n^\prime} $, and (iii) $\psi_{n}+\bar{\psi}_{n^\prime}\to \gamma$. In Ref.~\cite{Ghosh:2024hbf}, these rates were derived from the fermion self-energy by using two different methods. The final result is quoted in Eq.~(\ref{Gamma_n_pz-short}) in the main text, which is given in terms of function ${\cal M}_{n,n^{\prime}}(\xi)$ defined in Eq.~(\ref{Mnnp}). The latter is claimed to be proportional to the amplitude squared of the corresponding one-to-two and two-to-one processes. In this Appendix, we support this claim by calculating the squared amplitude explicitly. 
 
 Let us start from considering the following one-to-two process:  $\psi_n(P)\rightarrow \psi_{n^{\prime}}(P^{\prime})+\gamma(K)$. (The amplitudes squared for the other two processes are the same.) The $S$-matrix element for the process with an electron in the initial Landau level $n$ and the final Landau level $n^{\prime}$
 reads as
 \begin{equation}
  S_{fi} = g \int d^4 u \left< e(P^{\prime},n^{\prime})\gamma(K) \left| \bar{\psi} \gamma_\mu A^\mu \psi \right| e(P,n)\right> ,
 \end{equation}
 where $u^\mu\equiv (t,x,y,z)$ is the space-time coordinate.
 Using the solutions of the Dirac equation presented in Appendix~\ref{dirac_eq}, we can write the fermion field operator as follows:
  \begin{widetext}
  \begin{eqnarray}
  \psi(u) & =& \sum_{s=\pm} \sum_{n=0}^\infty \int \frac{dp_y  dp_z}{ (2\pi)^2\sqrt{2E_n}}\Big[ f_{s} (n, p_y,p_z) e^{-i p_\parallel \cdot u_\parallel - is_\perp p_y y } U_{s} (x_{p},n, p_z) +  
  \hat{f}_{s}^\dagger (n, p_y,p_z) e^{i p_\parallel \cdot u_\parallel + is_\perp p_y y } V_{s}(\tilde{x}_p,n, p_z) \Big] , 
  \label{psi}
 \end{eqnarray}
 where $f_{s}(n, p_y,p_z)$ is the annihilation operator for a fermion and $\hat{f}_{s}^\dagger(n, p_y,p_z)$ is the creation operator for an antifermion  in the $n$th Landau-level state with quantum numbers $p_y$ and $p_z$. The expressions for spinors $U_{s} (x_{p},n, p_z) $ and $V_{s}(\tilde{x}_p,n, p_z) $ in the Landau-level states with positive and negative energies are given in Appendix~\ref{dirac_eq}. The one-fermion states are defined by
 \begin{equation}
  \left| n, p_y,p_z \right> = \sqrt{2E_n}f_{s}^\dagger (n, p_y,p_z) \left| 0 \right> . 
 \end{equation}
 The Dirac conjugate of the fermion field is given by
 \begin{eqnarray}
  \bar{\psi}(u) &=& \sum_{s=\pm} \sum_{n=0}^\infty \int \frac{dp_y  dp_z}{   (2\pi)^2\sqrt{2E_n}}   \Big[ f_{s}^\dagger (n, p_y,p_z) e^{i p_\parallel \cdot u_\parallel + is_\perp p_y y}\bar{U}_{s} (x_{p},n, p_z) +  
  \hat{f}_{s} (n, p_y,p_z) e^{-i p_\parallel \cdot u_\parallel - is_\perp p_y y } \bar{V}_{s} (\tilde{x}_p,n, p_z) \Big] .
  \label{psi-bar}
 \end{eqnarray}
  The creation and annihilation operators satisfy the anticommutation relation
 \begin{eqnarray}
  \left\{ f_{s} (n, p_y,p_z), f_{s^{\prime}}^\dagger (n^{\prime}, p_y^{\prime},p_z^{\prime})
  \right\} &=& (2\pi)^2
  \delta_{ss^{\prime}} \delta_{nn^{\prime}} \delta(p_y-p^{\prime}_y) \delta (p_z - p^{\prime}_z) ,
  \label{freln}
 \end{eqnarray}
  \end{widetext}
 the operators $\hat{f}$ and $\hat{f}^\dagger$ satisfy a similar relation, while all other anticommutation relations give zeros. The equal time anticommutation relation for the fermion field operators reads
 \begin{equation}
  \left\{ \psi(u), \psi^\dagger(u^{\prime}) \right\} = \delta^3 (\bm{u} - \bm{u}^{\prime}) ,
  \label{equation}
 \end{equation}
 assuming $t=t^\prime$. The photon field operator takes the conventional form, 
 \begin{equation}
  A^\mu(u)=\int \frac{d^3\bm{k}}{(2\pi)^3\sqrt{2\omega_k}}\sum_r\epsilon^\mu_r(a_r e^{-ik\cdot u}+a_r^\dagger e^{ik\cdot u}),
 \end{equation}
 where $\epsilon^\mu_r$ are photon polarization vectors.
 
 By making use of the definitions of the field operators and the one-particle states, we derive  
 \begin{eqnarray}
  S_{fi} =  (2\pi)^3 \delta^2\left(p_{\parallel}^{\prime}+k_{\parallel}-p_{\parallel}\right)\delta\left(p_{ y}^{\prime}+k_{ y}-p_{ y}\right) 
  {M}_{ij}   ,
  \label{matrix-element}
 \end{eqnarray}
 where 
 \begin{equation}
  {M}_{ij}=g\int dx e^{ik_x x} \bar{U}_{s^{\prime}}(x_p^{\prime},n^{\prime},p_z^{\prime})\gamma^\mu \epsilon^r_\mu U_{s}(x_p,n,p_z) .
 \end{equation}
 Averaging over the initial spin states and summing over all final states (i.e., photon polarizations $r$ and fermion spin states $s^{\prime}$), we obtain
  \begin{widetext}
  \begin{eqnarray}
\sum_{r,s,s^\prime} \frac{ |{M}_{ij}|^2 }{\beta_n} &=& -\frac{g^2}{\beta_n} \int dx\int dx^{\prime} e^{ik_x(x-x^{\prime})} \mbox{tr}\left[P_U(x_{p^{\prime}}^{\prime},x_{p^{\prime}},n^{\prime},p_z^{\prime})\gamma^\mu P_U(x_p,x_p^{\prime},n,p_z)\gamma_\mu\right]\nonumber\\
  &=& -\frac{4g^2}{\beta_n} \int dx\int dx^{\prime} e^{ik_x(x-x^{\prime})}\bigg[m_{f}^2\bigg\{\phi_{n}(x_{p})\phi_{n}(x_{p}^{\prime})\phi_{n^{\prime}}(x_{p^{\prime}})\phi_{n^{\prime}}(x_{p^{\prime}}^{\prime})
  +\phi_{n-1}(x_{p})\phi_{n-1}(x_{p}^{\prime})\phi_{n^{\prime}-1}(x_{p^{\prime}})\phi_{n^{\prime}-1}(x_{p^{\prime}}^{\prime})\bigg\}\nonumber\\
  &&+(m_{f}^2-p_0 p_0^{\prime}+p_z p_z^{\prime})\bigg\{\phi_{n}(x_{p})\phi_{n}(x_{p}^{\prime})\phi_{n^{\prime}-1}(x_{p^{\prime}})\phi_{n^{\prime}-1}(x_{p^{\prime}}^{\prime})
  +\phi_{n-1}(x_{p})\phi_{n-1}(x_{p}^{\prime})\phi_{n^{\prime}}(x_{p^{\prime}})\phi_{n^{\prime}}(x_{p^{\prime}}^{\prime})\bigg\} \nonumber\\
  &&+2\sqrt{n |e_{f}B|}\sqrt{n^{\prime}|e_{f}B|}\bigg\{\phi_{n}(x_{p})\phi_{n-1}(x_{p}^{\prime})\phi_{n^{\prime}}(x_{p^{\prime}})\phi_{n^{\prime}-1}(x_{p^{\prime}}^{\prime})
  +\phi_{n-1}(x_{p})\phi_{n}(x_{p}^{\prime})\phi_{n^{\prime}-1}(x_{p^{\prime}})\phi_{n^{\prime}}(x_{p^{\prime}}^{\prime})\bigg\}\bigg] ,
  \label{Mij-squared}
 \end{eqnarray}
 where we used the Landau-level spin degeneracy factor $\beta_n\equiv 2-\delta_{n,0}$ and took into account that $\sum_r (\epsilon^r_\mu)^{*} \epsilon^r_\nu = - g_{\mu\nu}$.  The definition of the bi-spinor functions $P_U(x_p,x_p^{\prime},n,p_z)$ and the orbital wave functions $\phi_{n}(x_{p})$ are given in Appendix~\ref{dirac_eq}.  To integrate over $x$ and $x^{\prime}$, we use the following result: 
 \begin{eqnarray}
\int_{-\infty}^{\infty} dx\phi_n(x_p) \phi_{n^{\prime}}(x_{p^{\prime}})e^{i k_x x}  = \bigg( {2^n (n^{\prime})!  \over 2^{n^{\prime}} n! } \bigg)^{1/2} 
  e^{-\frac{1}{2}\zeta+\frac{i}{2}k_x(p_y+p_y^{\prime})\ell_{f}^2} 
  \bigg(\frac{\ell_{f}}{2}(p_y-p_y^{\prime}-ik_x)\bigg)^{n-n^{\prime}} L_{n^{\prime}}^{n-n^{\prime}}(\zeta) ,
  \label{int-phi-phi}
 \end{eqnarray}
 where $\zeta=\frac{1}{\beta_n}[k_x^2+(p_y-p_y^{\prime})^2]\ell_{f}^2$. Taking into account that $p_y-p_y^{\prime}=k_y$ due to the $\delta$-function in Eq.~(\ref{matrix-element}), we can replace $\zeta$ with $\xi= (k_\perp\ell_{f})^2/2$ in the amplitude squared.  Then, result in Eq.~(\ref{Mij-squared}) reduces down to
 \begin{eqnarray}
\sum_{r,s,s^\prime} \frac{ |{M}_{ij}|^2 }{\beta_n}  &=&-\frac{2g^2}{\beta_n}  \bigg\{m_{f}^2\ell_{f}^2 \left[ \mathcal{I}_0^{n,n^{\prime}}(\xi)+\mathcal{I}_0^{n-1,n^{\prime}-1}(\xi) \right]
   - \ell_{f}^2(m_{f}^2-p_0 p_0^{\prime}+p_z p_z^{\prime})\left[ \mathcal{I}_0^{n,n^{\prime}-1}(\xi)+\mathcal{I}_0^{n-1,n^{\prime}}(\xi)\right]
  -2  \mathcal{I}_2^{n-1,n^{\prime}-1}(\xi)\bigg\}\nonumber\\
  &=&\frac{2g^2}{\beta_n} \bigg\{-(n+n^{\prime}+m_{f}^2\ell_{f}^2)\left[\mathcal{I}_0^{n,n^{\prime}}\left(\xi\right)+\mathcal{I}_0^{n-1,n^{\prime}-1}\left(\xi\right)\right]
  +(n+n^{\prime})\left[\mathcal{I}_0^{n,n^{\prime}-1}\left(\xi\right)+\mathcal{I}_0^{n-1,n^{\prime}}\left(\xi\right)\right]\bigg\}.
  \label{Mij-squared-final}
 \end{eqnarray}
  \end{widetext}
As we see, the squared amplitude is proportional to ${\cal M}_{n,n^{\prime}}(\xi)/\beta_n$, which is indeed the function that appears in the definition of the damping rate in  Eq.~(\ref{Gamma_n_pz-short}). To obtain the final expression, we used the following relation:
 \begin{equation}
  p_0p_0^{\prime}-p_z p_z^{\prime} =(n+n^{\prime}-\xi)|e_{f}B|+m_{f}^2, 
 \end{equation}
 which follows from the energy-momentum conservation \cite{Ghosh:2024hbf}. We also used the following functions~\cite{Wang:2021ebh}: 
 \begin{eqnarray}
  \mathcal{I}_{0}^{n,n^{\prime}}(\xi)&=&
  \frac{(n^\prime)!}{n!} e^{-\xi}  \xi^{n-n^\prime} \left(L_{n^\prime}^{n-n^\prime}\left(\xi\right)\right)^2 \nonumber\\
&=& \frac{n!}{(n^\prime)!}e^{-\xi} \xi^{n^\prime-n} \left(L_{n}^{n^\prime-n}\left(\xi\right)\right)^2 ,
  \label{I0f-LL-form1}  
 \end{eqnarray} 
 \begin{eqnarray}
  \mathcal{I}_{2}^{n,n^{\prime}}(\xi)  &=& 
  \frac{n+n^{\prime}+2}{2}\left[\mathcal{I}_{0}^{n,n^{\prime}}(\xi) +\mathcal{I}_{0}^{n+1,n^{\prime}+1}(\xi) \right]
   \nonumber\\
&&  -\frac{\xi}{2}\left[\mathcal{I}_{0}^{n+1,n^{\prime}}(\xi) +\mathcal{I}_{0}^{n,n^{\prime}+1}(\xi) \right] .
  \label{I2-I0}
 \end{eqnarray}

 \section{Dirac equation in a uniform magnetic field}
 \label{dirac_eq}
 
 The Dirac equation in the presence of a magnetic field is given by
 \begin{equation}
  \left( i\gamma^\mu D_{\mu} - m_{f}\right)\psi (u) =0,
  \label{dirac}
 \end{equation}
 where the covariant derivative is $D_{\mu} = \partial_{\mu} + ie_{f} A_{\mu}$ and the background gauge field is $\bm{A}=(0,Bx,0)$. By using  
 the Dirac representation of $\gamma$ matrices:
 \begin{equation}
  \gamma^0 =\left(\begin{array}{ll}
   \mathbb{I}_2 & 0\\
   0 & - \mathbb{I}_2
  \end{array}\right), \qquad
  \gamma^i =\left(\begin{array}{ll}
   0 & \sigma_i \\
   -\sigma_i & 0 
  \end{array}\right),
 \end{equation}
 where $\sigma_i$ are the Pauli matrices, we derive the following positive-energy solutions of the Dirac equation:
 \begin{equation}
  \Phi^{+}_{n,p,s}(u)=\frac{1}{\sqrt{2\pi}} e^{-i p_\parallel \cdot u_\parallel - is_\perp p_y y }U_{s} (x_{p},n, p_z) ,
 \end{equation}
 with the pair of $n$th Landau-level spinors $U_{s}(x_{p},n, p_z)$ given by
 \begin{eqnarray}
  U_{+} (x_{p},n, p_z) &=& \sqrt{E_n+m_f}\left( \begin{array}{c} 
   \phi_{n}(x_p) \\
   0 \\[1ex]
   \frac{p_z}{E_n+m_f} \phi_{n}(x_p) \\[1ex]
   -i\frac{\sqrt{2n|e_{f}B|}}{E_n+m_f} \phi_{n-1} (x_p) 
  \end{array} \right) , \nonumber \\
\end{eqnarray}
\begin{eqnarray}
  U_{-} (x_{p},n, p_z) &=&\sqrt{E_n+m_f}\left( \begin{array}{c} 
   0 \\
   i\phi_{n-1}(x_p) \\[1ex]
   -\frac{\sqrt{2n|e_{f}B|}}{E_n+m_f} \phi_{n}(x_p) \\[1ex]
   -i\frac{p_z }{E_n+m_f} \phi_{n-1} (x_p) 
  \end{array} \right) . 
  \label{Usoln}\nonumber\\
 \end{eqnarray}
 Here $x_p=x/\ell_{f}+p_y \ell_{f}$, $\ell_{f}=1/\sqrt{|e_{f}B|}$ is the magnetic length, and $E_n =\sqrt{2n|e_{f}B|+p_z^2+m_f^2}$ is the Landau-level energy. In the last two expressions, we assumed that $s_{\perp}=\mbox{sign}(e_{f}B)>0$. The expressions for the spinors are similar when $s_\perp=-1$, but functions $\phi_{n} (x_p)$ and $i\phi_{n-1} (x_p)$ switch places. Similar spinors (for $s_\perp=-1$) were found in Ref.~\cite{Bhattacharya:2002qf}.
 The orbital wave functions have the following explicit expressions
 \begin{equation}
  \phi_{n}( x_p)= \frac{1}{\sqrt{2^n n!\sqrt{\pi}\ell_{f} }} H_n\left(x_p\right)e^{-x_p^2/2} ,
 \end{equation}
 where $H_{n}(x)$ are the Hermite polynomials.
 
  When calculating the amplitudes squared of the leading-order processes in Appendix~\ref{app:amplitude}, one uses the following bi-spinor functions, obtained by summing the Landau-level spinors over the spins of the initial (final) states,
\begin{widetext} 
\begin{eqnarray}
  P_U(x_p,x_p^{\prime},n,p_z)&=& \sum_{s} U_{s} (x_p,n,p_z)  \bar{U}_{s} (x_p^\prime,n,p_z)
  = \frac{1 }{2}
  \Big\{ \left(E_{n}\gamma^0-p_z\gamma^3+m_{f}\right)   \left[\phi_{n}(x_p)\phi_{n} (x_p^\prime)+\phi_{n-1}(x_p)\phi_{n-1} (x_p^\prime) \right] \nonumber\\
  &&+i \gamma^1 \gamma^2 \left(E_{n}\gamma^0-p_z\gamma^3+m_{f}\right)  \left[\phi_{n}(x_p)\phi_{n} (x_p^\prime)-\phi_{n-1}(x_p)\phi_{n-1} (x_p^\prime) \right]  \nonumber\\
  &&+\sqrt{2n|e_{f}B|} (i \gamma^1 +\gamma^2) \phi_{n-1}(x_p)\phi_{n} (x_p^\prime)
  +\sqrt{2n|e_{f}B|} (-i \gamma^1 +\gamma^2)  \phi_{n}(x_p)\phi_{n-1} (x_p^\prime) \Big\}.
 \end{eqnarray}
This is analogous to the result obtained in Ref.~\cite{Bhattacharya:2002qf}.
 Following the same approach, one can also derive the negative energy spinors, 
 \begin{equation}
  \Phi^{-}_{n,p,s}(u)=\frac{1}{\sqrt{2\pi}} e^{i p_\parallel \cdot u_\parallel + is_\perp p_{y} y } V_{s} (\tilde{x}_{p},n, p_z) ,
 \end{equation}
 where $\tilde{x}_p = x/\ell_{f} - p_y \ell_{f}$ and 
 \begin{eqnarray}
  V_{+} (\tilde{x}_p,n, p_z) &=& \sqrt{E_n+m_f}\left( \begin{array}{c} 
   \frac{p_z }{E_n+m_f} \phi_{n}(\tilde{x}_p) \\[1ex]
   i\frac{\sqrt{2n|e_{f}B|}}{E_n+m_f} \phi_{n-1} (\tilde{x}_p) \\[1ex]
   \phi_{n}(\tilde{x}_p) \\ 
   0 
  \end{array} \right) , \nonumber\\
\\  
  V_{-} (\tilde{x}_p,n, p_z) &=& \sqrt{E_n+m_f}\left( \begin{array}{c} 
   \frac{\sqrt{2n|e_{f}B|}}{E_n+m_f} \phi_{n}(\tilde{x}_p) \\[1ex]
   -i\frac{p_z}{E_n+m_f} \phi_{n-1} (\tilde{x}_p) \\[1ex]
   0 \\ 
   i\phi_{n-1}(\tilde{x}_p)  
  \end{array} \right) , \nonumber\\
  \label{Vsoln}
 \end{eqnarray}
 assuming $s_{\perp}=\mbox{sign}(e_{f}B)>0$. When $s_\perp=-1$, functions $\phi_{n} (\tilde{x}_p)$ and $i\phi_{n-1} (\tilde{x}_p)$ switch places in the definition  of $V_{s} (\tilde{x}_{p},n, p_z) $. The bi-spinor function obtained from the negative-energy spinors reads as
 \begin{eqnarray}
  P_V(\tilde{x}_p,\tilde{x}_p^{\prime},n,p_z)
  &=&\sum_{s} V_{s} (\tilde{x}_p,n,p_z)  \bar{V}_{s} (\tilde{x}_p^\prime,n,p_z) 
 = \frac{1}{2} \Big\{ \left(E_{n}\gamma^0-p_z\gamma^3-m_{f}\right) \left[\phi_{n}(\tilde{x}_p)\phi_{n} (\tilde{x}_p^\prime)+\phi_{n-1}(\tilde{x}_p)\phi_{n-1} (\tilde{x}_p^\prime) \right]  \nonumber\\
  &&+i \gamma^1 \gamma^2\left(E_{n}\gamma^0-p_z\gamma^3-m_{f}\right)  \left[\phi_{n}(\tilde{x}_p)\phi_{n} (\tilde{x}_p^\prime)-\phi_{n-1}(\tilde{x}_p)\phi_{n-1} (\tilde{x}_p^\prime) \right]  \nonumber\\
  &&-\sqrt{2n|e_{f}B|} (i \gamma^1 +\gamma^2) \phi_{n-1}(\tilde{x}_p)\phi_{n} (\tilde{x}_p^\prime)
  -\sqrt{2n|e_{f}B|} (-i \gamma^1 +\gamma^2)  \phi_{n}(\tilde{x}_p)\phi_{n-1} (\tilde{x}_p^\prime) \Big\}.
 \end{eqnarray}
  \end{widetext} 


\begin{thebibliography}{66}%
\makeatletter
\providecommand \@ifxundefined [1]{%
 \@ifx{#1\undefined}
}%
\providecommand \@ifnum [1]{%
 \ifnum #1\expandafter \@firstoftwo
 \else \expandafter \@secondoftwo
 \fi
}%
\providecommand \@ifx [1]{%
 \ifx #1\expandafter \@firstoftwo
 \else \expandafter \@secondoftwo
 \fi
}%
\providecommand \natexlab [1]{#1}%
\providecommand \enquote  [1]{``#1''}%
\providecommand \bibnamefont  [1]{#1}%
\providecommand \bibfnamefont [1]{#1}%
\providecommand \citenamefont [1]{#1}%
\providecommand \href@noop [0]{\@secondoftwo}%
\providecommand \href [0]{\begingroup \@sanitize@url \@href}%
\providecommand \@href[1]{\@@startlink{#1}\@@href}%
\providecommand \@@href[1]{\endgroup#1\@@endlink}%
\providecommand \@sanitize@url [0]{\catcode `\\12\catcode `\$12\catcode
  `\&12\catcode `\#12\catcode `\^12\catcode `\_12\catcode `\%12\relax}%
\providecommand \@@startlink[1]{}%
\providecommand \@@endlink[0]{}%
\providecommand \url  [0]{\begingroup\@sanitize@url \@url }%
\providecommand \@url [1]{\endgroup\@href {#1}{\urlprefix }}%
\providecommand \urlprefix  [0]{URL }%
\providecommand \Eprint [0]{\href }%
\providecommand \doibase [0]{http://dx.doi.org/}%
\providecommand \selectlanguage [0]{\@gobble}%
\providecommand \bibinfo  [0]{\@secondoftwo}%
\providecommand \bibfield  [0]{\@secondoftwo}%
\providecommand \translation [1]{[#1]}%
\providecommand \BibitemOpen [0]{}%
\providecommand \bibitemStop [0]{}%
\providecommand \bibitemNoStop [0]{.\EOS\space}%
\providecommand \EOS [0]{\spacefactor3000\relax}%
\providecommand \BibitemShut  [1]{\csname bibitem#1\endcsname}%
\let\auto@bib@innerbib\@empty
\bibitem [{\citenamefont {Sarri}\ \emph {et~al.}(2015)\citenamefont {Sarri}
  \emph {et~al.}}]{Sarri:2015jyr}%
  \BibitemOpen
  \bibfield  {author} {\bibinfo {author} {\bibfnamefont {G.}~\bibnamefont
  {Sarri}} \emph {et~al.},\ }\href {\doibase 10.1038/ncomms7747} {\bibfield
  {journal} {\bibinfo  {journal} {Nature Commun.}\ }\textbf {\bibinfo {volume}
  {6}},\ \bibinfo {pages} {6747} (\bibinfo {year} {2015})}\BibitemShut
  {NoStop}%
\bibitem [{\citenamefont {Chen}\ and\ \citenamefont {Fiuza}(2023)}]{Chen_2023}%
  \BibitemOpen
  \bibfield  {author} {\bibinfo {author} {\bibfnamefont {H.}~\bibnamefont
  {Chen}}\ and\ \bibinfo {author} {\bibfnamefont {F.}~\bibnamefont {Fiuza}},\
  }\href {\doibase 10.1063/5.0134819} {\bibfield  {journal} {\bibinfo
  {journal} {Physics of Plasmas}\ }\textbf {\bibinfo {volume} {30}},\ \bibinfo
  {pages} {020601} (\bibinfo {year} {2023})}\BibitemShut {NoStop}%
\bibitem [{\citenamefont {Sturrock}(1971)}]{Sturrock:1971zc}%
  \BibitemOpen
  \bibfield  {author} {\bibinfo {author} {\bibfnamefont {P.~A.}\ \bibnamefont
  {Sturrock}},\ }\href {\doibase 10.1086/150865} {\bibfield  {journal}
  {\bibinfo  {journal} {Astrophys. J.}\ }\textbf {\bibinfo {volume} {164}},\
  \bibinfo {pages} {529} (\bibinfo {year} {1971})}\BibitemShut {NoStop}%
\bibitem [{\citenamefont {Ruderman}\ and\ \citenamefont
  {Sutherland}(1975)}]{Ruderman:1975ju}%
  \BibitemOpen
  \bibfield  {author} {\bibinfo {author} {\bibfnamefont {M.~A.}\ \bibnamefont
  {Ruderman}}\ and\ \bibinfo {author} {\bibfnamefont {P.~G.}\ \bibnamefont
  {Sutherland}},\ }\href {\doibase 10.1086/153393} {\bibfield  {journal}
  {\bibinfo  {journal} {Astrophys. J.}\ }\textbf {\bibinfo {volume} {196}},\
  \bibinfo {pages} {51} (\bibinfo {year} {1975})}\BibitemShut {NoStop}%
\bibitem [{\citenamefont {Arons}(1983)}]{Arons:1983aa}%
  \BibitemOpen
  \bibfield  {author} {\bibinfo {author} {\bibfnamefont {J.}~\bibnamefont
  {Arons}},\ }\href@noop {} {\bibfield  {journal} {\bibinfo  {journal}
  {Astrophys. J.}\ }\textbf {\bibinfo {volume} {266}},\ \bibinfo {pages} {215}
  (\bibinfo {year} {1983})}\BibitemShut {NoStop}%
\bibitem [{\citenamefont {Turolla}\ \emph {et~al.}(2015)\citenamefont
  {Turolla}, \citenamefont {Zane},\ and\ \citenamefont
  {Watts}}]{Turolla:2015mwa}%
  \BibitemOpen
  \bibfield  {author} {\bibinfo {author} {\bibfnamefont {R.}~\bibnamefont
  {Turolla}}, \bibinfo {author} {\bibfnamefont {S.}~\bibnamefont {Zane}}, \
  and\ \bibinfo {author} {\bibfnamefont {A.}~\bibnamefont {Watts}},\ }\href
  {\doibase 10.1088/0034-4885/78/11/116901} {\bibfield  {journal} {\bibinfo
  {journal} {Rep. Prog. Phys.}\ }\textbf {\bibinfo {volume} {78}},\ \bibinfo
  {pages} {116901} (\bibinfo {year} {2015})},\ \Eprint
  {http://arxiv.org/abs/1507.02924} {arXiv:1507.02924 [astro-ph.HE]}
  \BibitemShut {NoStop}%
\bibitem [{\citenamefont {Kaspi}\ and\ \citenamefont
  {Beloborodov}(2017)}]{Kaspi:2017fwg}%
  \BibitemOpen
  \bibfield  {author} {\bibinfo {author} {\bibfnamefont {V.~M.}\ \bibnamefont
  {Kaspi}}\ and\ \bibinfo {author} {\bibfnamefont {A.}~\bibnamefont
  {Beloborodov}},\ }\href {\doibase 10.1146/annurev-astro-081915-023329}
  {\bibfield  {journal} {\bibinfo  {journal} {Ann. Rev. Astron. Astrophys.}\
  }\textbf {\bibinfo {volume} {55}},\ \bibinfo {pages} {261} (\bibinfo {year}
  {2017})},\ \Eprint {http://arxiv.org/abs/1703.00068} {arXiv:1703.00068
  [astro-ph.HE]} \BibitemShut {NoStop}%
\bibitem [{\citenamefont {Grasso}\ and\ \citenamefont
  {Rubinstein}(2001)}]{Grasso:2000wj}%
  \BibitemOpen
  \bibfield  {author} {\bibinfo {author} {\bibfnamefont {D.}~\bibnamefont
  {Grasso}}\ and\ \bibinfo {author} {\bibfnamefont {H.~R.}\ \bibnamefont
  {Rubinstein}},\ }\href {\doibase 10.1016/S0370-1573(00)00110-1} {\bibfield
  {journal} {\bibinfo  {journal} {Phys. Rep.}\ }\textbf {\bibinfo {volume}
  {348}},\ \bibinfo {pages} {163} (\bibinfo {year} {2001})},\ \Eprint
  {http://arxiv.org/abs/astro-ph/0009061} {arXiv:astro-ph/0009061} \BibitemShut
  {NoStop}%
\bibitem [{\citenamefont {Yagi}\ \emph {et~al.}(2005)\citenamefont {Yagi},
  \citenamefont {Hatsuda},\ and\ \citenamefont {Miake}}]{Yagi:2005yb}%
  \BibitemOpen
  \bibfield  {author} {\bibinfo {author} {\bibfnamefont {K.}~\bibnamefont
  {Yagi}}, \bibinfo {author} {\bibfnamefont {T.}~\bibnamefont {Hatsuda}}, \
  and\ \bibinfo {author} {\bibfnamefont {Y.}~\bibnamefont {Miake}},\
  }\href@noop {} {\emph {\bibinfo {title} {{Quark-Gluon Plasma: From Big Bang
  to Little Bang}}}},\ Vol.~\bibinfo {volume} {23}\ (\bibinfo  {publisher}
  {Cambridge University Press},\ \bibinfo {address} {Cambridge, UK},\ \bibinfo
  {year} {2005})\BibitemShut {NoStop}%
\bibitem [{\citenamefont {Adams}\ \emph {et~al.}(2005)\citenamefont {Adams}
  \emph {et~al.}}]{STAR:2005gfr}%
  \BibitemOpen
  \bibfield  {author} {\bibinfo {author} {\bibfnamefont {J.}~\bibnamefont
  {Adams}} \emph {et~al.} (\bibinfo {collaboration} {STAR}),\ }\href {\doibase
  10.1016/j.nuclphysa.2005.03.085} {\bibfield  {journal} {\bibinfo  {journal}
  {Nucl. Phys. A}\ }\textbf {\bibinfo {volume} {757}},\ \bibinfo {pages} {102}
  (\bibinfo {year} {2005})},\ \Eprint {http://arxiv.org/abs/nucl-ex/0501009}
  {arXiv:nucl-ex/0501009} \BibitemShut {NoStop}%
\bibitem [{\citenamefont {Adcox}\ \emph {et~al.}(2005)\citenamefont {Adcox}
  \emph {et~al.}}]{PHENIX:2004vcz}%
  \BibitemOpen
  \bibfield  {author} {\bibinfo {author} {\bibfnamefont {K.}~\bibnamefont
  {Adcox}} \emph {et~al.} (\bibinfo {collaboration} {PHENIX}),\ }\href
  {\doibase 10.1016/j.nuclphysa.2005.03.086} {\bibfield  {journal} {\bibinfo
  {journal} {Nucl. Phys. A}\ }\textbf {\bibinfo {volume} {757}},\ \bibinfo
  {pages} {184} (\bibinfo {year} {2005})},\ \Eprint
  {http://arxiv.org/abs/nucl-ex/0410003} {arXiv:nucl-ex/0410003} \BibitemShut
  {NoStop}%
\bibitem [{\citenamefont {Back}\ \emph {et~al.}(2005)\citenamefont {Back} \emph
  {et~al.}}]{PHOBOS:2004zne}%
  \BibitemOpen
  \bibfield  {author} {\bibinfo {author} {\bibfnamefont {B.~B.}\ \bibnamefont
  {Back}} \emph {et~al.} (\bibinfo {collaboration} {PHOBOS}),\ }\href {\doibase
  10.1016/j.nuclphysa.2005.03.084} {\bibfield  {journal} {\bibinfo  {journal}
  {Nucl. Phys. A}\ }\textbf {\bibinfo {volume} {757}},\ \bibinfo {pages} {28}
  (\bibinfo {year} {2005})},\ \Eprint {http://arxiv.org/abs/nucl-ex/0410022}
  {arXiv:nucl-ex/0410022} \BibitemShut {NoStop}%
\bibitem [{\citenamefont {Durrer}\ and\ \citenamefont
  {Neronov}(2013)}]{Durrer:2013pga}%
  \BibitemOpen
  \bibfield  {author} {\bibinfo {author} {\bibfnamefont {R.}~\bibnamefont
  {Durrer}}\ and\ \bibinfo {author} {\bibfnamefont {A.}~\bibnamefont
  {Neronov}},\ }\href {\doibase 10.1007/s00159-013-0062-7} {\bibfield
  {journal} {\bibinfo  {journal} {Astron. Astrophys. Rev.}\ }\textbf {\bibinfo
  {volume} {21}},\ \bibinfo {pages} {62} (\bibinfo {year} {2013})},\ \Eprint
  {http://arxiv.org/abs/1303.7121} {arXiv:1303.7121} \BibitemShut {NoStop}%
\bibitem [{\citenamefont {Vachaspati}(2021)}]{Vachaspati:2020blt}%
  \BibitemOpen
  \bibfield  {author} {\bibinfo {author} {\bibfnamefont {T.}~\bibnamefont
  {Vachaspati}},\ }\href {\doibase 10.1088/1361-6633/ac03a9} {\bibfield
  {journal} {\bibinfo  {journal} {Rep. Prog. Phys.}\ }\textbf {\bibinfo
  {volume} {84}},\ \bibinfo {pages} {074901} (\bibinfo {year} {2021})},\
  \Eprint {http://arxiv.org/abs/2010.10525} {arXiv:2010.10525 [astro-ph.CO]}
  \BibitemShut {NoStop}%
\bibitem [{\citenamefont {Skokov}\ \emph {et~al.}(2009)\citenamefont {Skokov},
  \citenamefont {Illarionov},\ and\ \citenamefont {Toneev}}]{Skokov:2009qp}%
  \BibitemOpen
  \bibfield  {author} {\bibinfo {author} {\bibfnamefont {V.}~\bibnamefont
  {Skokov}}, \bibinfo {author} {\bibfnamefont {A.~Y.}\ \bibnamefont
  {Illarionov}}, \ and\ \bibinfo {author} {\bibfnamefont {V.}~\bibnamefont
  {Toneev}},\ }\href {\doibase 10.1142/S0217751X09047570} {\bibfield  {journal}
  {\bibinfo  {journal} {Int. J. Mod. Phys. A}\ }\textbf {\bibinfo {volume}
  {24}},\ \bibinfo {pages} {5925} (\bibinfo {year} {2009})},\ \Eprint
  {http://arxiv.org/abs/0907.1396} {arXiv:0907.1396 [nucl-th]} \BibitemShut
  {NoStop}%
\bibitem [{\citenamefont {Voronyuk}\ \emph {et~al.}(2011)\citenamefont
  {Voronyuk}, \citenamefont {Toneev}, \citenamefont {Cassing}, \citenamefont
  {Bratkovskaya}, \citenamefont {Konchakovski},\ and\ \citenamefont
  {Voloshin}}]{Voronyuk:2011jd}%
  \BibitemOpen
  \bibfield  {author} {\bibinfo {author} {\bibfnamefont {V.}~\bibnamefont
  {Voronyuk}}, \bibinfo {author} {\bibfnamefont {V.~D.}\ \bibnamefont
  {Toneev}}, \bibinfo {author} {\bibfnamefont {W.}~\bibnamefont {Cassing}},
  \bibinfo {author} {\bibfnamefont {E.~L.}\ \bibnamefont {Bratkovskaya}},
  \bibinfo {author} {\bibfnamefont {V.~P.}\ \bibnamefont {Konchakovski}}, \
  and\ \bibinfo {author} {\bibfnamefont {S.~A.}\ \bibnamefont {Voloshin}},\
  }\href {\doibase 10.1103/PhysRevC.83.054911} {\bibfield  {journal} {\bibinfo
  {journal} {Phys. Rev. C}\ }\textbf {\bibinfo {volume} {83}},\ \bibinfo
  {pages} {054911} (\bibinfo {year} {2011})},\ \Eprint
  {http://arxiv.org/abs/1103.4239} {arXiv:1103.4239 [nucl-th]} \BibitemShut
  {NoStop}%
\bibitem [{\citenamefont {Deng}\ and\ \citenamefont
  {Huang}(2012)}]{Deng:2012pc}%
  \BibitemOpen
  \bibfield  {author} {\bibinfo {author} {\bibfnamefont {W.-T.}\ \bibnamefont
  {Deng}}\ and\ \bibinfo {author} {\bibfnamefont {X.-G.}\ \bibnamefont
  {Huang}},\ }\href {\doibase 10.1103/PhysRevC.85.044907} {\bibfield  {journal}
  {\bibinfo  {journal} {Phys. Rev. C}\ }\textbf {\bibinfo {volume} {85}},\
  \bibinfo {pages} {044907} (\bibinfo {year} {2012})},\ \Eprint
  {http://arxiv.org/abs/1201.5108} {arXiv:1201.5108 [nucl-th]} \BibitemShut
  {NoStop}%
\bibitem [{\citenamefont {Bloczynski}\ \emph {et~al.}(2013)\citenamefont
  {Bloczynski}, \citenamefont {Huang}, \citenamefont {Zhang},\ and\
  \citenamefont {Liao}}]{Bloczynski:2012en}%
  \BibitemOpen
  \bibfield  {author} {\bibinfo {author} {\bibfnamefont {J.}~\bibnamefont
  {Bloczynski}}, \bibinfo {author} {\bibfnamefont {X.-G.}\ \bibnamefont
  {Huang}}, \bibinfo {author} {\bibfnamefont {X.}~\bibnamefont {Zhang}}, \ and\
  \bibinfo {author} {\bibfnamefont {J.}~\bibnamefont {Liao}},\ }\href {\doibase
  10.1016/j.physletb.2012.12.030} {\bibfield  {journal} {\bibinfo  {journal}
  {Phys. Lett. B}\ }\textbf {\bibinfo {volume} {718}},\ \bibinfo {pages} {1529}
  (\bibinfo {year} {2013})},\ \Eprint {http://arxiv.org/abs/1209.6594}
  {arXiv:1209.6594 [nucl-th]} \BibitemShut {NoStop}%
\bibitem [{\citenamefont {Guo}\ \emph {et~al.}(2020)\citenamefont {Guo},
  \citenamefont {Liao},\ and\ \citenamefont {Wang}}]{Guo:2019mgh}%
  \BibitemOpen
  \bibfield  {author} {\bibinfo {author} {\bibfnamefont {X.}~\bibnamefont
  {Guo}}, \bibinfo {author} {\bibfnamefont {J.}~\bibnamefont {Liao}}, \ and\
  \bibinfo {author} {\bibfnamefont {E.}~\bibnamefont {Wang}},\ }\href {\doibase
  10.1038/s41598-020-59129-6} {\bibfield  {journal} {\bibinfo  {journal} {Sci.
  Rep.}\ }\textbf {\bibinfo {volume} {10}},\ \bibinfo {pages} {2196} (\bibinfo
  {year} {2020})},\ \Eprint {http://arxiv.org/abs/1904.04704} {arXiv:1904.04704
  [hep-ph]} \BibitemShut {NoStop}%
\bibitem [{\citenamefont {{van Erkelens}}\ and\ \citenamefont {{van
  Leeuwen}}(1984)}]{vanErkelens:1984}%
  \BibitemOpen
  \bibfield  {author} {\bibinfo {author} {\bibfnamefont {H.}~\bibnamefont {{van
  Erkelens}}}\ and\ \bibinfo {author} {\bibfnamefont {W.}~\bibnamefont {{van
  Leeuwen}}},\ }\href {\doibase https://doi.org/10.1016/0378-4371(84)90104-3}
  {\bibfield  {journal} {\bibinfo  {journal} {Physica A}\ }\textbf {\bibinfo
  {volume} {123}},\ \bibinfo {pages} {72} (\bibinfo {year} {1984})}\BibitemShut
  {NoStop}%
\bibitem [{\citenamefont {Pike}\ and\ \citenamefont
  {Rose}(2016)}]{Pike:2016aa}%
  \BibitemOpen
  \bibfield  {author} {\bibinfo {author} {\bibfnamefont {O.~J.}\ \bibnamefont
  {Pike}}\ and\ \bibinfo {author} {\bibfnamefont {S.~J.}\ \bibnamefont
  {Rose}},\ }\href {\doibase 10.1103/PhysRevE.93.053208} {\bibfield  {journal}
  {\bibinfo  {journal} {Phys. Rev. E}\ }\textbf {\bibinfo {volume} {93}},\
  \bibinfo {pages} {053208} (\bibinfo {year} {2016})}\BibitemShut {NoStop}%
\bibitem [{\citenamefont {Hattori}\ and\ \citenamefont
  {Satow}(2016)}]{Hattori:2016cnt}%
  \BibitemOpen
  \bibfield  {author} {\bibinfo {author} {\bibfnamefont {K.}~\bibnamefont
  {Hattori}}\ and\ \bibinfo {author} {\bibfnamefont {D.}~\bibnamefont
  {Satow}},\ }\href {\doibase 10.1103/PhysRevD.94.114032} {\bibfield  {journal}
  {\bibinfo  {journal} {Phys. Rev. D}\ }\textbf {\bibinfo {volume} {94}},\
  \bibinfo {pages} {114032} (\bibinfo {year} {2016})},\ \Eprint
  {http://arxiv.org/abs/1610.06818} {arXiv:1610.06818 [hep-ph]} \BibitemShut
  {NoStop}%
\bibitem [{\citenamefont {Hattori}\ \emph {et~al.}(2017)\citenamefont
  {Hattori}, \citenamefont {Li}, \citenamefont {Satow},\ and\ \citenamefont
  {Yee}}]{Hattori:2016lqx}%
  \BibitemOpen
  \bibfield  {author} {\bibinfo {author} {\bibfnamefont {K.}~\bibnamefont
  {Hattori}}, \bibinfo {author} {\bibfnamefont {S.}~\bibnamefont {Li}},
  \bibinfo {author} {\bibfnamefont {D.}~\bibnamefont {Satow}}, \ and\ \bibinfo
  {author} {\bibfnamefont {H.-U.}\ \bibnamefont {Yee}},\ }\href {\doibase
  10.1103/PhysRevD.95.076008} {\bibfield  {journal} {\bibinfo  {journal} {Phys.
  Rev. D}\ }\textbf {\bibinfo {volume} {95}},\ \bibinfo {pages} {076008}
  (\bibinfo {year} {2017})},\ \Eprint {http://arxiv.org/abs/1610.06839}
  {arXiv:1610.06839 [hep-ph]} \BibitemShut {NoStop}%
\bibitem [{\citenamefont {Fukushima}\ and\ \citenamefont
  {Hidaka}(2018)}]{Fukushima:2017lvb}%
  \BibitemOpen
  \bibfield  {author} {\bibinfo {author} {\bibfnamefont {K.}~\bibnamefont
  {Fukushima}}\ and\ \bibinfo {author} {\bibfnamefont {Y.}~\bibnamefont
  {Hidaka}},\ }\href {\doibase 10.1103/PhysRevLett.120.162301} {\bibfield
  {journal} {\bibinfo  {journal} {Phys. Rev. Lett.}\ }\textbf {\bibinfo
  {volume} {120}},\ \bibinfo {pages} {162301} (\bibinfo {year} {2018})},\
  \Eprint {http://arxiv.org/abs/1711.01472} {arXiv:1711.01472 [hep-ph]}
  \BibitemShut {NoStop}%
\bibitem [{\citenamefont {Fukushima}\ and\ \citenamefont
  {Hidaka}(2020)}]{Fukushima:2019ugr}%
  \BibitemOpen
  \bibfield  {author} {\bibinfo {author} {\bibfnamefont {K.}~\bibnamefont
  {Fukushima}}\ and\ \bibinfo {author} {\bibfnamefont {Y.}~\bibnamefont
  {Hidaka}},\ }\href {\doibase 10.1007/JHEP04(2020)162} {\bibfield  {journal}
  {\bibinfo  {journal} {J. High Energy Phys.}\ }\textbf {\bibinfo {volume}
  {04}},\ \bibinfo {pages} {162} (\bibinfo {year} {2020})},\ \Eprint
  {http://arxiv.org/abs/1906.02683} {arXiv:1906.02683 [hep-ph]} \BibitemShut
  {NoStop}%
\bibitem [{\citenamefont {Buividovich}\ and\ \citenamefont
  {Polikarpov}(2011)}]{Buividovich:2010qe}%
  \BibitemOpen
  \bibfield  {author} {\bibinfo {author} {\bibfnamefont {P.~V.}\ \bibnamefont
  {Buividovich}}\ and\ \bibinfo {author} {\bibfnamefont {M.~I.}\ \bibnamefont
  {Polikarpov}},\ }\href {\doibase 10.1103/PhysRevD.83.094508} {\bibfield
  {journal} {\bibinfo  {journal} {Phys. Rev. D}\ }\textbf {\bibinfo {volume}
  {83}},\ \bibinfo {pages} {094508} (\bibinfo {year} {2011})},\ \Eprint
  {http://arxiv.org/abs/1011.3001} {arXiv:1011.3001 [hep-lat]} \BibitemShut
  {NoStop}%
\bibitem [{\citenamefont {Buividovich}\ \emph {et~al.}(2010)\citenamefont
  {Buividovich}, \citenamefont {Chernodub}, \citenamefont {Kharzeev},
  \citenamefont {Kalaydzhyan}, \citenamefont {Luschevskaya},\ and\
  \citenamefont {Polikarpov}}]{Buividovich:2010tn}%
  \BibitemOpen
  \bibfield  {author} {\bibinfo {author} {\bibfnamefont {P.~V.}\ \bibnamefont
  {Buividovich}}, \bibinfo {author} {\bibfnamefont {M.~N.}\ \bibnamefont
  {Chernodub}}, \bibinfo {author} {\bibfnamefont {D.~E.}\ \bibnamefont
  {Kharzeev}}, \bibinfo {author} {\bibfnamefont {T.}~\bibnamefont
  {Kalaydzhyan}}, \bibinfo {author} {\bibfnamefont {E.~V.}\ \bibnamefont
  {Luschevskaya}}, \ and\ \bibinfo {author} {\bibfnamefont {M.~I.}\
  \bibnamefont {Polikarpov}},\ }\href {\doibase 10.1103/PhysRevLett.105.132001}
  {\bibfield  {journal} {\bibinfo  {journal} {Phys. Rev. Lett.}\ }\textbf
  {\bibinfo {volume} {105}},\ \bibinfo {pages} {132001} (\bibinfo {year}
  {2010})},\ \Eprint {http://arxiv.org/abs/1003.2180} {arXiv:1003.2180
  [hep-lat]} \BibitemShut {NoStop}%
\bibitem [{\citenamefont {Astrakhantsev}\ \emph {et~al.}(2020)\citenamefont
  {Astrakhantsev}, \citenamefont {Braguta}, \citenamefont {D'Elia},
  \citenamefont {Kotov}, \citenamefont {Nikolaev},\ and\ \citenamefont
  {Sanfilippo}}]{Astrakhantsev:2019zkr}%
  \BibitemOpen
  \bibfield  {author} {\bibinfo {author} {\bibfnamefont {N.}~\bibnamefont
  {Astrakhantsev}}, \bibinfo {author} {\bibfnamefont {V.~V.}\ \bibnamefont
  {Braguta}}, \bibinfo {author} {\bibfnamefont {M.}~\bibnamefont {D'Elia}},
  \bibinfo {author} {\bibfnamefont {A.~Y.}\ \bibnamefont {Kotov}}, \bibinfo
  {author} {\bibfnamefont {A.~A.}\ \bibnamefont {Nikolaev}}, \ and\ \bibinfo
  {author} {\bibfnamefont {F.}~\bibnamefont {Sanfilippo}},\ }\href {\doibase
  10.1103/PhysRevD.102.054516} {\bibfield  {journal} {\bibinfo  {journal}
  {Phys. Rev. D}\ }\textbf {\bibinfo {volume} {102}},\ \bibinfo {pages}
  {054516} (\bibinfo {year} {2020})},\ \Eprint
  {http://arxiv.org/abs/1910.08516} {arXiv:1910.08516 [hep-lat]} \BibitemShut
  {NoStop}%
\bibitem [{\citenamefont {Almirante}\ \emph {et~al.}(2024)\citenamefont
  {Almirante}, \citenamefont {Astrakhantsev}, \citenamefont {Braguta},
  \citenamefont {D'Elia}, \citenamefont {Maio}, \citenamefont {Naviglio},
  \citenamefont {Sanfilippo},\ and\ \citenamefont
  {Trunin}}]{Almirante:2024lqn}%
  \BibitemOpen
  \bibfield  {author} {\bibinfo {author} {\bibfnamefont {G.}~\bibnamefont
  {Almirante}}, \bibinfo {author} {\bibfnamefont {N.}~\bibnamefont
  {Astrakhantsev}}, \bibinfo {author} {\bibfnamefont {V.~V.}\ \bibnamefont
  {Braguta}}, \bibinfo {author} {\bibfnamefont {M.}~\bibnamefont {D'Elia}},
  \bibinfo {author} {\bibfnamefont {L.}~\bibnamefont {Maio}}, \bibinfo {author}
  {\bibfnamefont {M.}~\bibnamefont {Naviglio}}, \bibinfo {author}
  {\bibfnamefont {F.}~\bibnamefont {Sanfilippo}}, \ and\ \bibinfo {author}
  {\bibfnamefont {A.}~\bibnamefont {Trunin}},\ }\href@noop {} {\  (\bibinfo
  {year} {2024})},\ \Eprint {http://arxiv.org/abs/2406.18504} {arXiv:2406.18504
  [hep-lat]} \BibitemShut {NoStop}%
\bibitem [{\citenamefont {Mamo}(2013)}]{Mamo:2012kqw}%
  \BibitemOpen
  \bibfield  {author} {\bibinfo {author} {\bibfnamefont {K.~A.}\ \bibnamefont
  {Mamo}},\ }\href {\doibase 10.1007/JHEP08(2013)083} {\bibfield  {journal}
  {\bibinfo  {journal} {J. High Energy Phys.}\ }\textbf {\bibinfo {volume}
  {08}},\ \bibinfo {pages} {083} (\bibinfo {year} {2013})},\ \Eprint
  {http://arxiv.org/abs/1210.7428} {arXiv:1210.7428 [hep-th]} \BibitemShut
  {NoStop}%
\bibitem [{\citenamefont {Fukushima}\ and\ \citenamefont
  {Okutsu}(2022)}]{Fukushima:2021got}%
  \BibitemOpen
  \bibfield  {author} {\bibinfo {author} {\bibfnamefont {K.}~\bibnamefont
  {Fukushima}}\ and\ \bibinfo {author} {\bibfnamefont {A.}~\bibnamefont
  {Okutsu}},\ }\href {\doibase 10.1103/PhysRevD.105.054016} {\bibfield
  {journal} {\bibinfo  {journal} {Phys. Rev. D}\ }\textbf {\bibinfo {volume}
  {105}},\ \bibinfo {pages} {054016} (\bibinfo {year} {2022})},\ \Eprint
  {http://arxiv.org/abs/2106.07968} {arXiv:2106.07968 [hep-ph]} \BibitemShut
  {NoStop}%
\bibitem [{\citenamefont {Nam}(2012)}]{Nam:2012sg}%
  \BibitemOpen
  \bibfield  {author} {\bibinfo {author} {\bibfnamefont {S.-i.}\ \bibnamefont
  {Nam}},\ }\href {\doibase 10.1103/PhysRevD.86.033014} {\bibfield  {journal}
  {\bibinfo  {journal} {Phys. Rev. D}\ }\textbf {\bibinfo {volume} {86}},\
  \bibinfo {pages} {033014} (\bibinfo {year} {2012})},\ \Eprint
  {http://arxiv.org/abs/1207.3172} {arXiv:1207.3172 [hep-ph]} \BibitemShut
  {NoStop}%
\bibitem [{\citenamefont {Kerbikov}\ and\ \citenamefont
  {Andreichikov}(2015)}]{Kerbikov:2014ofa}%
  \BibitemOpen
  \bibfield  {author} {\bibinfo {author} {\bibfnamefont {B.~O.}\ \bibnamefont
  {Kerbikov}}\ and\ \bibinfo {author} {\bibfnamefont {M.~A.}\ \bibnamefont
  {Andreichikov}},\ }\href {\doibase 10.1103/PhysRevD.91.074010} {\bibfield
  {journal} {\bibinfo  {journal} {Phys. Rev. D}\ }\textbf {\bibinfo {volume}
  {91}},\ \bibinfo {pages} {074010} (\bibinfo {year} {2015})},\ \Eprint
  {http://arxiv.org/abs/1410.3413} {arXiv:1410.3413 [hep-ph]} \BibitemShut
  {NoStop}%
\bibitem [{\citenamefont {Satapathy}\ \emph {et~al.}(2021)\citenamefont
  {Satapathy}, \citenamefont {Ghosh},\ and\ \citenamefont
  {Ghosh}}]{Satapathy:2021cjp}%
  \BibitemOpen
  \bibfield  {author} {\bibinfo {author} {\bibfnamefont {S.}~\bibnamefont
  {Satapathy}}, \bibinfo {author} {\bibfnamefont {S.}~\bibnamefont {Ghosh}}, \
  and\ \bibinfo {author} {\bibfnamefont {S.}~\bibnamefont {Ghosh}},\ }\href
  {\doibase 10.1103/PhysRevD.104.056030} {\bibfield  {journal} {\bibinfo
  {journal} {Phys. Rev. D}\ }\textbf {\bibinfo {volume} {104}},\ \bibinfo
  {pages} {056030} (\bibinfo {year} {2021})},\ \Eprint
  {http://arxiv.org/abs/2104.03917} {arXiv:2104.03917 [hep-ph]} \BibitemShut
  {NoStop}%
\bibitem [{\citenamefont {Satapathy}\ \emph {et~al.}(2022)\citenamefont
  {Satapathy}, \citenamefont {Ghosh},\ and\ \citenamefont
  {Ghosh}}]{Satapathy:2021wex}%
  \BibitemOpen
  \bibfield  {author} {\bibinfo {author} {\bibfnamefont {S.}~\bibnamefont
  {Satapathy}}, \bibinfo {author} {\bibfnamefont {S.}~\bibnamefont {Ghosh}}, \
  and\ \bibinfo {author} {\bibfnamefont {S.}~\bibnamefont {Ghosh}},\ }\href
  {\doibase 10.1103/PhysRevD.106.036006} {\bibfield  {journal} {\bibinfo
  {journal} {Phys. Rev. D}\ }\textbf {\bibinfo {volume} {106}},\ \bibinfo
  {pages} {036006} (\bibinfo {year} {2022})},\ \Eprint
  {http://arxiv.org/abs/2112.08236} {arXiv:2112.08236 [hep-ph]} \BibitemShut
  {NoStop}%
\bibitem [{\citenamefont {Bandyopadhyay}\ \emph {et~al.}(2023)\citenamefont
  {Bandyopadhyay}, \citenamefont {Ghosh}, \citenamefont {Farias},\ and\
  \citenamefont {Ghosh}}]{Bandyopadhyay:2023lvk}%
  \BibitemOpen
  \bibfield  {author} {\bibinfo {author} {\bibfnamefont {A.}~\bibnamefont
  {Bandyopadhyay}}, \bibinfo {author} {\bibfnamefont {S.}~\bibnamefont
  {Ghosh}}, \bibinfo {author} {\bibfnamefont {R.~L.~S.}\ \bibnamefont
  {Farias}}, \ and\ \bibinfo {author} {\bibfnamefont {S.}~\bibnamefont
  {Ghosh}},\ }\href {\doibase 10.1140/epjc/s10052-023-11655-z} {\bibfield
  {journal} {\bibinfo  {journal} {Eur. Phys. J. C}\ }\textbf {\bibinfo {volume}
  {83}},\ \bibinfo {pages} {489} (\bibinfo {year} {2023})},\ \Eprint
  {http://arxiv.org/abs/2305.15844} {arXiv:2305.15844 [hep-ph]} \BibitemShut
  {NoStop}%
\bibitem [{\citenamefont {Kurian}\ and\ \citenamefont
  {Chandra}(2017)}]{Kurian:2017yxj}%
  \BibitemOpen
  \bibfield  {author} {\bibinfo {author} {\bibfnamefont {M.}~\bibnamefont
  {Kurian}}\ and\ \bibinfo {author} {\bibfnamefont {V.}~\bibnamefont
  {Chandra}},\ }\href {\doibase 10.1103/PhysRevD.96.114026} {\bibfield
  {journal} {\bibinfo  {journal} {Phys. Rev. D}\ }\textbf {\bibinfo {volume}
  {96}},\ \bibinfo {pages} {114026} (\bibinfo {year} {2017})},\ \Eprint
  {http://arxiv.org/abs/1709.08320} {arXiv:1709.08320 [nucl-th]} \BibitemShut
  {NoStop}%
\bibitem [{\citenamefont {Das}\ \emph {et~al.}(2020)\citenamefont {Das},
  \citenamefont {Mishra},\ and\ \citenamefont {Mohapatra}}]{Das:2019ppb}%
  \BibitemOpen
  \bibfield  {author} {\bibinfo {author} {\bibfnamefont {A.}~\bibnamefont
  {Das}}, \bibinfo {author} {\bibfnamefont {H.}~\bibnamefont {Mishra}}, \ and\
  \bibinfo {author} {\bibfnamefont {R.~K.}\ \bibnamefont {Mohapatra}},\ }\href
  {\doibase 10.1103/PhysRevD.101.034027} {\bibfield  {journal} {\bibinfo
  {journal} {Phys. Rev. D}\ }\textbf {\bibinfo {volume} {101}},\ \bibinfo
  {pages} {034027} (\bibinfo {year} {2020})},\ \Eprint
  {http://arxiv.org/abs/1907.05298} {arXiv:1907.05298 [hep-ph]} \BibitemShut
  {NoStop}%
\bibitem [{\citenamefont {Thakur}\ and\ \citenamefont
  {Srivastava}(2019)}]{Thakur:2019bnf}%
  \BibitemOpen
  \bibfield  {author} {\bibinfo {author} {\bibfnamefont {L.}~\bibnamefont
  {Thakur}}\ and\ \bibinfo {author} {\bibfnamefont {P.~K.}\ \bibnamefont
  {Srivastava}},\ }\href {\doibase 10.1103/PhysRevD.100.076016} {\bibfield
  {journal} {\bibinfo  {journal} {Phys. Rev. D}\ }\textbf {\bibinfo {volume}
  {100}},\ \bibinfo {pages} {076016} (\bibinfo {year} {2019})},\ \Eprint
  {http://arxiv.org/abs/1910.12087} {arXiv:1910.12087 [hep-ph]} \BibitemShut
  {NoStop}%
\bibitem [{\citenamefont {Dey}\ \emph {et~al.}(2022)\citenamefont {Dey},
  \citenamefont {Samanta}, \citenamefont {Ghosh},\ and\ \citenamefont
  {Satapathy}}]{Dey:2020awu}%
  \BibitemOpen
  \bibfield  {author} {\bibinfo {author} {\bibfnamefont {J.}~\bibnamefont
  {Dey}}, \bibinfo {author} {\bibfnamefont {S.}~\bibnamefont {Samanta}},
  \bibinfo {author} {\bibfnamefont {S.}~\bibnamefont {Ghosh}}, \ and\ \bibinfo
  {author} {\bibfnamefont {S.}~\bibnamefont {Satapathy}},\ }\href {\doibase
  10.1103/PhysRevC.106.044914} {\bibfield  {journal} {\bibinfo  {journal}
  {Phys. Rev. C}\ }\textbf {\bibinfo {volume} {106}},\ \bibinfo {pages}
  {044914} (\bibinfo {year} {2022})},\ \Eprint
  {http://arxiv.org/abs/2002.04434} {arXiv:2002.04434 [nucl-th]} \BibitemShut
  {NoStop}%
\bibitem [{\citenamefont {Wang}\ and\ \citenamefont
  {Shovkovy}(2021)}]{Wang:2021ebh}%
  \BibitemOpen
  \bibfield  {author} {\bibinfo {author} {\bibfnamefont {X.}~\bibnamefont
  {Wang}}\ and\ \bibinfo {author} {\bibfnamefont {I.}~\bibnamefont
  {Shovkovy}},\ }\href {\doibase 10.1103/PhysRevD.104.056017} {\bibfield
  {journal} {\bibinfo  {journal} {Phys. Rev. D}\ }\textbf {\bibinfo {volume}
  {104}},\ \bibinfo {pages} {056017} (\bibinfo {year} {2021})},\ \Eprint
  {http://arxiv.org/abs/2103.01967} {arXiv:2103.01967 [nucl-th]} \BibitemShut
  {NoStop}%
\bibitem [{\citenamefont {Ghosh}\ and\ \citenamefont
  {Shovkovy}(2024{\natexlab{a}})}]{Ghosh:2024hbf}%
  \BibitemOpen
  \bibfield  {author} {\bibinfo {author} {\bibfnamefont {R.}~\bibnamefont
  {Ghosh}}\ and\ \bibinfo {author} {\bibfnamefont {I.~A.}\ \bibnamefont
  {Shovkovy}},\ }\href {\doibase 10.1103/PhysRevD.109.096018} {\bibfield
  {journal} {\bibinfo  {journal} {Phys. Rev. D}\ }\textbf {\bibinfo {volume}
  {109}},\ \bibinfo {pages} {096018} (\bibinfo {year} {2024}{\natexlab{a}})},\
  \Eprint {http://arxiv.org/abs/2402.04307} {arXiv:2402.04307 [hep-ph]}
  \BibitemShut {NoStop}%
\bibitem [{\citenamefont {Ghosh}\ and\ \citenamefont
  {Shovkovy}(2024{\natexlab{b}})}]{Ghosh:2024fkg}%
  \BibitemOpen
  \bibfield  {author} {\bibinfo {author} {\bibfnamefont {R.}~\bibnamefont
  {Ghosh}}\ and\ \bibinfo {author} {\bibfnamefont {I.~A.}\ \bibnamefont
  {Shovkovy}},\ }\href {\doibase 10.1103/PhysRevD.110.096009} {\bibfield
  {journal} {\bibinfo  {journal} {Phys. Rev. D}\ }\textbf {\bibinfo {volume}
  {110}},\ \bibinfo {pages} {096009} (\bibinfo {year} {2024}{\natexlab{b}})},\
  \Eprint {http://arxiv.org/abs/2404.01388} {arXiv:2404.01388 [hep-ph]}
  \BibitemShut {NoStop}%
\bibitem [{\citenamefont {Miransky}\ and\ \citenamefont
  {Shovkovy}(2015)}]{Miransky:2015ava}%
  \BibitemOpen
  \bibfield  {author} {\bibinfo {author} {\bibfnamefont {V.~A.}\ \bibnamefont
  {Miransky}}\ and\ \bibinfo {author} {\bibfnamefont {I.~A.}\ \bibnamefont
  {Shovkovy}},\ }\href {\doibase 10.1016/j.physrep.2015.02.003} {\bibfield
  {journal} {\bibinfo  {journal} {Phys. Rep.}\ }\textbf {\bibinfo {volume}
  {576}},\ \bibinfo {pages} {1} (\bibinfo {year} {2015})},\ \Eprint
  {http://arxiv.org/abs/1503.00732} {arXiv:1503.00732} \BibitemShut {NoStop}%
\bibitem [{\citenamefont {Aarts}\ and\ \citenamefont
  {Martinez~Resco}(2002)}]{Aarts:2002tn}%
  \BibitemOpen
  \bibfield  {author} {\bibinfo {author} {\bibfnamefont {G.}~\bibnamefont
  {Aarts}}\ and\ \bibinfo {author} {\bibfnamefont {J.~M.}\ \bibnamefont
  {Martinez~Resco}},\ }\href {\doibase 10.1088/1126-6708/2002/11/022}
  {\bibfield  {journal} {\bibinfo  {journal} {JHEP}\ }\textbf {\bibinfo
  {volume} {11}},\ \bibinfo {pages} {022} (\bibinfo {year} {2002})},\ \Eprint
  {http://arxiv.org/abs/hep-ph/0209048} {arXiv:hep-ph/0209048} \BibitemShut
  {NoStop}%
\bibitem [{\citenamefont {Gradshteyn}\ and\ \citenamefont
  {Ryzhik}(1980)}]{Gradshteyn:1943cpj}%
  \BibitemOpen
  \bibfield  {author} {\bibinfo {author} {\bibfnamefont {I.~S.}\ \bibnamefont
  {Gradshteyn}}\ and\ \bibinfo {author} {\bibfnamefont {I.~M.}\ \bibnamefont
  {Ryzhik}},\ }\href@noop {} {\emph {\bibinfo {title} {{Table of Integrals,
  Series, and Products}}}},\ \bibinfo {edition} {5th}\ ed.\ (\bibinfo
  {publisher} {Academic Press},\ \bibinfo {address} {New York},\ \bibinfo
  {year} {1980})\BibitemShut {NoStop}%
\bibitem [{\citenamefont {Weldon}(1983)}]{Weldon:1983jn}%
  \BibitemOpen
  \bibfield  {author} {\bibinfo {author} {\bibfnamefont {H.~A.}\ \bibnamefont
  {Weldon}},\ }\href {\doibase 10.1103/PhysRevD.28.2007} {\bibfield  {journal}
  {\bibinfo  {journal} {Phys. Rev. D}\ }\textbf {\bibinfo {volume} {28}},\
  \bibinfo {pages} {2007} (\bibinfo {year} {1983})}\BibitemShut {NoStop}%
\bibitem [{Note1()}]{Note1}%
  \BibitemOpen
  \bibinfo {note} {The proof that the squared amplitude of the leading-order
  processes is proportional to function ${\protect \cal M}_{n,n^{\prime }} (\xi
  )$ is given in Appendix~\ref {app:amplitude}.}\BibitemShut {Stop}%
\bibitem [{QED()}]{QED-QCDCond:2024}%
  \BibitemOpen
  \href@noop {} {\enquote {\bibinfo {title} {{See Supplemental Material {\em
  for numerical data for the electrical conductivity in strongly magnetized QED
  and QCD plasmas}}},}\ }\bibinfo {howpublished}
  {\url{https://www.dropbox.com/scl/fo/7tmcx67idkl5mo4k6k307/APVt4p-MWBYdArWYcmz0dkQ?rlkey=un1vl1jrabzof0outfvz62pc6&dl=0}}\BibitemShut
  {NoStop}%
\bibitem [{\citenamefont {McLerran}\ and\ \citenamefont
  {Skokov}(2014)}]{McLerran:2013hla}%
  \BibitemOpen
  \bibfield  {author} {\bibinfo {author} {\bibfnamefont {L.}~\bibnamefont
  {McLerran}}\ and\ \bibinfo {author} {\bibfnamefont {V.}~\bibnamefont
  {Skokov}},\ }\href {\doibase 10.1016/j.nuclphysa.2014.05.008} {\bibfield
  {journal} {\bibinfo  {journal} {Nucl. Phys. A}\ }\textbf {\bibinfo {volume}
  {929}},\ \bibinfo {pages} {184} (\bibinfo {year} {2014})},\ \Eprint
  {http://arxiv.org/abs/1305.0774} {arXiv:1305.0774 [hep-ph]} \BibitemShut
  {NoStop}%
\bibitem [{\citenamefont {Tuchin}(2016)}]{Tuchin:2015oka}%
  \BibitemOpen
  \bibfield  {author} {\bibinfo {author} {\bibfnamefont {K.}~\bibnamefont
  {Tuchin}},\ }\href {\doibase 10.1103/PhysRevC.93.014905} {\bibfield
  {journal} {\bibinfo  {journal} {Phys. Rev. C}\ }\textbf {\bibinfo {volume}
  {93}},\ \bibinfo {pages} {014905} (\bibinfo {year} {2016})},\ \Eprint
  {http://arxiv.org/abs/1508.06925} {arXiv:1508.06925 [hep-ph]} \BibitemShut
  {NoStop}%
\bibitem [{\citenamefont {Yan}\ and\ \citenamefont
  {Huang}(2023)}]{Yan:2021zjc}%
  \BibitemOpen
  \bibfield  {author} {\bibinfo {author} {\bibfnamefont {L.}~\bibnamefont
  {Yan}}\ and\ \bibinfo {author} {\bibfnamefont {X.-G.}\ \bibnamefont
  {Huang}},\ }\href {\doibase 10.1103/PhysRevD.107.094028} {\bibfield
  {journal} {\bibinfo  {journal} {Phys. Rev. D}\ }\textbf {\bibinfo {volume}
  {107}},\ \bibinfo {pages} {094028} (\bibinfo {year} {2023})},\ \Eprint
  {http://arxiv.org/abs/2104.00831} {arXiv:2104.00831 [nucl-th]} \BibitemShut
  {NoStop}%
\bibitem [{\citenamefont {Kim}\ \emph {et~al.}(2013)\citenamefont {Kim},
  \citenamefont {Kim}, \citenamefont {Wang}, \citenamefont {Sasaki},
  \citenamefont {Satoh}, \citenamefont {Ohnishi}, \citenamefont {Kitaura},
  \citenamefont {Yang},\ and\ \citenamefont {Li}}]{Kim:2013ys}%
  \BibitemOpen
  \bibfield  {author} {\bibinfo {author} {\bibfnamefont {H.-J.}\ \bibnamefont
  {Kim}}, \bibinfo {author} {\bibfnamefont {K.-S.}\ \bibnamefont {Kim}},
  \bibinfo {author} {\bibfnamefont {J.~F.}\ \bibnamefont {Wang}}, \bibinfo
  {author} {\bibfnamefont {M.}~\bibnamefont {Sasaki}}, \bibinfo {author}
  {\bibfnamefont {N.}~\bibnamefont {Satoh}}, \bibinfo {author} {\bibfnamefont
  {A.}~\bibnamefont {Ohnishi}}, \bibinfo {author} {\bibfnamefont
  {M.}~\bibnamefont {Kitaura}}, \bibinfo {author} {\bibfnamefont
  {M.}~\bibnamefont {Yang}}, \ and\ \bibinfo {author} {\bibfnamefont
  {L.}~\bibnamefont {Li}},\ }\href {http://arxiv.org/abs/1307.6990} {\bibfield
  {journal} {\bibinfo  {journal} {Phys. Rev. Lett.}\ }\textbf {\bibinfo
  {volume} {111}},\ \bibinfo {pages} {246603} (\bibinfo {year} {2013})},\
  \Eprint {http://arxiv.org/abs/1307.6990} {arXiv:1307.6990} \BibitemShut
  {NoStop}%
\bibitem [{\citenamefont {Li}\ \emph {et~al.}(2016{\natexlab{a}})\citenamefont
  {Li}, \citenamefont {Kharzeev}, \citenamefont {Zhang}, \citenamefont {Huang},
  \citenamefont {Pletikosic}, \citenamefont {Fedorov}, \citenamefont {Zhong},
  \citenamefont {Schneeloch}, \citenamefont {Gu},\ and\ \citenamefont
  {Valla}}]{Li:2014bha}%
  \BibitemOpen
  \bibfield  {author} {\bibinfo {author} {\bibfnamefont {Q.}~\bibnamefont
  {Li}}, \bibinfo {author} {\bibfnamefont {D.~E.}\ \bibnamefont {Kharzeev}},
  \bibinfo {author} {\bibfnamefont {C.}~\bibnamefont {Zhang}}, \bibinfo
  {author} {\bibfnamefont {Y.}~\bibnamefont {Huang}}, \bibinfo {author}
  {\bibfnamefont {I.}~\bibnamefont {Pletikosic}}, \bibinfo {author}
  {\bibfnamefont {A.~V.}\ \bibnamefont {Fedorov}}, \bibinfo {author}
  {\bibfnamefont {R.~D.}\ \bibnamefont {Zhong}}, \bibinfo {author}
  {\bibfnamefont {J.~A.}\ \bibnamefont {Schneeloch}}, \bibinfo {author}
  {\bibfnamefont {G.~D.}\ \bibnamefont {Gu}}, \ and\ \bibinfo {author}
  {\bibfnamefont {T.}~\bibnamefont {Valla}},\ }\href {\doibase
  10.1038/nphys3648} {\bibfield  {journal} {\bibinfo  {journal} {Nature Phys.}\
  }\textbf {\bibinfo {volume} {12}},\ \bibinfo {pages} {550} (\bibinfo {year}
  {2016}{\natexlab{a}})},\ \Eprint {http://arxiv.org/abs/1412.6543}
  {arXiv:1412.6543} \BibitemShut {NoStop}%
\bibitem [{\citenamefont {{Xiong}}\ \emph {et~al.}(2015)\citenamefont
  {{Xiong}}, \citenamefont {{Kushwaha}}, \citenamefont {{Liang}}, \citenamefont
  {{Krizan}}, \citenamefont {{Hirschberger}}, \citenamefont {{Wang}},
  \citenamefont {{Cava}},\ and\ \citenamefont {{Ong}}}]{Xiong:2015yq}%
  \BibitemOpen
  \bibfield  {author} {\bibinfo {author} {\bibfnamefont {J.}~\bibnamefont
  {{Xiong}}}, \bibinfo {author} {\bibfnamefont {S.~K.}\ \bibnamefont
  {{Kushwaha}}}, \bibinfo {author} {\bibfnamefont {T.}~\bibnamefont {{Liang}}},
  \bibinfo {author} {\bibfnamefont {J.~W.}\ \bibnamefont {{Krizan}}}, \bibinfo
  {author} {\bibfnamefont {M.}~\bibnamefont {{Hirschberger}}}, \bibinfo
  {author} {\bibfnamefont {W.}~\bibnamefont {{Wang}}}, \bibinfo {author}
  {\bibfnamefont {R.~J.}\ \bibnamefont {{Cava}}}, \ and\ \bibinfo {author}
  {\bibfnamefont {N.~P.}\ \bibnamefont {{Ong}}},\ }\href {\doibase
  10.1126/science.aac6089} {\bibfield  {journal} {\bibinfo  {journal}
  {Science}\ }\textbf {\bibinfo {volume} {350}},\ \bibinfo {pages} {413}
  (\bibinfo {year} {2015})}\BibitemShut {NoStop}%
\bibitem [{\citenamefont {Feng}\ \emph {et~al.}(2015)\citenamefont {Feng},
  \citenamefont {Pang}, \citenamefont {Wu}, \citenamefont {Wang}, \citenamefont
  {Weng}, \citenamefont {Li}, \citenamefont {Dai}, \citenamefont {Fang},
  \citenamefont {Shi},\ and\ \citenamefont {Lu}}]{Feng:2015PRB92}%
  \BibitemOpen
  \bibfield  {author} {\bibinfo {author} {\bibfnamefont {J.}~\bibnamefont
  {Feng}}, \bibinfo {author} {\bibfnamefont {Y.}~\bibnamefont {Pang}}, \bibinfo
  {author} {\bibfnamefont {D.}~\bibnamefont {Wu}}, \bibinfo {author}
  {\bibfnamefont {Z.}~\bibnamefont {Wang}}, \bibinfo {author} {\bibfnamefont
  {H.}~\bibnamefont {Weng}}, \bibinfo {author} {\bibfnamefont {J.}~\bibnamefont
  {Li}}, \bibinfo {author} {\bibfnamefont {X.}~\bibnamefont {Dai}}, \bibinfo
  {author} {\bibfnamefont {Z.}~\bibnamefont {Fang}}, \bibinfo {author}
  {\bibfnamefont {Y.}~\bibnamefont {Shi}}, \ and\ \bibinfo {author}
  {\bibfnamefont {L.}~\bibnamefont {Lu}},\ }\href {\doibase
  10.1103/PhysRevB.92.081306} {\bibfield  {journal} {\bibinfo  {journal} {Phys.
  Rev. B}\ }\textbf {\bibinfo {volume} {92}},\ \bibinfo {pages} {081306}
  (\bibinfo {year} {2015})}\BibitemShut {NoStop}%
\bibitem [{\citenamefont {Li}\ \emph {et~al.}(2015)\citenamefont {Li},
  \citenamefont {Wang}, \citenamefont {Liu}, \citenamefont {Wang},
  \citenamefont {Liao},\ and\ \citenamefont {Yu}}]{Li-Yu-Cd3As2:2015}%
  \BibitemOpen
  \bibfield  {author} {\bibinfo {author} {\bibfnamefont {C.-Z.}\ \bibnamefont
  {Li}}, \bibinfo {author} {\bibfnamefont {L.-X.}\ \bibnamefont {Wang}},
  \bibinfo {author} {\bibfnamefont {H.}~\bibnamefont {Liu}}, \bibinfo {author}
  {\bibfnamefont {J.}~\bibnamefont {Wang}}, \bibinfo {author} {\bibfnamefont
  {Z.-M.}\ \bibnamefont {Liao}}, \ and\ \bibinfo {author} {\bibfnamefont
  {D.-P.}\ \bibnamefont {Yu}},\ }\href {\doibase 10.1038/ncomms10137}
  {\bibfield  {journal} {\bibinfo  {journal} {Nature Commun.}\ }\textbf
  {\bibinfo {volume} {6}},\ \bibinfo {pages} {10137} (\bibinfo {year}
  {2015})}\BibitemShut {NoStop}%
\bibitem [{\citenamefont {Li}\ \emph {et~al.}(2016{\natexlab{b}})\citenamefont
  {Li}, \citenamefont {He}, \citenamefont {Lu}, \citenamefont {Zhang},
  \citenamefont {Liu}, \citenamefont {Ma}, \citenamefont {Fan}, \citenamefont
  {Shen},\ and\ \citenamefont {Wang}}]{Li-Wang-Cd3As2:2016}%
  \BibitemOpen
  \bibfield  {author} {\bibinfo {author} {\bibfnamefont {H.}~\bibnamefont
  {Li}}, \bibinfo {author} {\bibfnamefont {H.}~\bibnamefont {He}}, \bibinfo
  {author} {\bibfnamefont {H.-Z.}\ \bibnamefont {Lu}}, \bibinfo {author}
  {\bibfnamefont {H.}~\bibnamefont {Zhang}}, \bibinfo {author} {\bibfnamefont
  {H.}~\bibnamefont {Liu}}, \bibinfo {author} {\bibfnamefont {R.}~\bibnamefont
  {Ma}}, \bibinfo {author} {\bibfnamefont {Z.}~\bibnamefont {Fan}}, \bibinfo
  {author} {\bibfnamefont {S.-Q.}\ \bibnamefont {Shen}}, \ and\ \bibinfo
  {author} {\bibfnamefont {J.}~\bibnamefont {Wang}},\ }\href {\doibase
  10.1038/ncomms10301} {\bibfield  {journal} {\bibinfo  {journal} {Nature
  Commun.}\ }\textbf {\bibinfo {volume} {7}},\ \bibinfo {pages} {10301}
  (\bibinfo {year} {2016}{\natexlab{b}})}\BibitemShut {NoStop}%
\bibitem [{\citenamefont {Huang}\ \emph {et~al.}(2015)\citenamefont {Huang}
  \emph {et~al.}}]{Huang:2015ve}%
  \BibitemOpen
  \bibfield  {author} {\bibinfo {author} {\bibfnamefont {X.}~\bibnamefont
  {Huang}} \emph {et~al.},\ }\href {https://arxiv.org/abs/1503.01304}
  {\bibfield  {journal} {\bibinfo  {journal} {Phys. Rev. X}\ }\textbf {\bibinfo
  {volume} {5}},\ \bibinfo {pages} {031023} (\bibinfo {year} {2015})},\ \Eprint
  {http://arxiv.org/abs/1503.01304} {arXiv:1503.01304} \BibitemShut {NoStop}%
\bibitem [{\citenamefont {Zhang}\ \emph {et~al.}(2016)\citenamefont {Zhang}
  \emph {et~al.}}]{Zhang:2015gwa}%
  \BibitemOpen
  \bibfield  {author} {\bibinfo {author} {\bibfnamefont {C.}~\bibnamefont
  {Zhang}} \emph {et~al.},\ }\href {https://doi.org/10.1038/ncomms10735}
  {\bibfield  {journal} {\bibinfo  {journal} {Nature Commun.}\ }\textbf
  {\bibinfo {volume} {7}},\ \bibinfo {pages} {0735} (\bibinfo {year} {2016})},\
  \Eprint {http://arxiv.org/abs/1601.04208} {arXiv:1601.04208} \BibitemShut
  {NoStop}%
\bibitem [{\citenamefont {Ambjorn}\ \emph {et~al.}(1983)\citenamefont
  {Ambjorn}, \citenamefont {Greensite},\ and\ \citenamefont
  {Peterson}}]{Ambjorn:1983hp}%
  \BibitemOpen
  \bibfield  {author} {\bibinfo {author} {\bibfnamefont {J.}~\bibnamefont
  {Ambjorn}}, \bibinfo {author} {\bibfnamefont {J.}~\bibnamefont {Greensite}},
  \ and\ \bibinfo {author} {\bibfnamefont {C.}~\bibnamefont {Peterson}},\
  }\href {\doibase 10.1016/0550-3213(83)90585-0} {\bibfield  {journal}
  {\bibinfo  {journal} {Nucl. Phys.}\ }\textbf {\bibinfo {volume} {B221}},\
  \bibinfo {pages} {381} (\bibinfo {year} {1983})}\BibitemShut {NoStop}%
\bibitem [{\citenamefont {Abdulhamid}\ \emph {et~al.}(2024)\citenamefont
  {Abdulhamid} \emph {et~al.}}]{STAR:2023jdd}%
  \BibitemOpen
  \bibfield  {author} {\bibinfo {author} {\bibfnamefont {M.~I.}\ \bibnamefont
  {Abdulhamid}} \emph {et~al.} (\bibinfo {collaboration} {STAR}),\ }\href
  {\doibase 10.1103/PhysRevX.14.011028} {\bibfield  {journal} {\bibinfo
  {journal} {Phys. Rev. X}\ }\textbf {\bibinfo {volume} {14}},\ \bibinfo
  {pages} {011028} (\bibinfo {year} {2024})},\ \Eprint
  {http://arxiv.org/abs/2304.03430} {arXiv:2304.03430 [nucl-ex]} \BibitemShut
  {NoStop}%
\bibitem [{\citenamefont {Li}\ \emph {et~al.}(2012)\citenamefont {Li},
  \citenamefont {Spitkovsky},\ and\ \citenamefont {Tchekhovskoy}}]{Li:2011zh}%
  \BibitemOpen
  \bibfield  {author} {\bibinfo {author} {\bibfnamefont {J.}~\bibnamefont
  {Li}}, \bibinfo {author} {\bibfnamefont {A.}~\bibnamefont {Spitkovsky}}, \
  and\ \bibinfo {author} {\bibfnamefont {A.}~\bibnamefont {Tchekhovskoy}},\
  }\href {\doibase 10.1088/0004-637X/746/1/60} {\bibfield  {journal} {\bibinfo
  {journal} {Astrophys. J.}\ }\textbf {\bibinfo {volume} {746}},\ \bibinfo
  {pages} {60} (\bibinfo {year} {2012})},\ \Eprint
  {http://arxiv.org/abs/1107.0979} {arXiv:1107.0979 [astro-ph.HE]} \BibitemShut
  {NoStop}%
\bibitem [{\citenamefont {Kalapotharakos}\ \emph
  {et~al.}(2012{\natexlab{a}})\citenamefont {Kalapotharakos}, \citenamefont
  {Kazanas}, \citenamefont {Harding},\ and\ \citenamefont
  {Contopoulos}}]{Kalapotharakos:2011vg}%
  \BibitemOpen
  \bibfield  {author} {\bibinfo {author} {\bibfnamefont {C.}~\bibnamefont
  {Kalapotharakos}}, \bibinfo {author} {\bibfnamefont {D.}~\bibnamefont
  {Kazanas}}, \bibinfo {author} {\bibfnamefont {A.}~\bibnamefont {Harding}}, \
  and\ \bibinfo {author} {\bibfnamefont {I.}~\bibnamefont {Contopoulos}},\
  }\href {\doibase 10.1088/0004-637X/749/1/2} {\bibfield  {journal} {\bibinfo
  {journal} {Astrophys. J.}\ }\textbf {\bibinfo {volume} {749}},\ \bibinfo
  {pages} {2} (\bibinfo {year} {2012}{\natexlab{a}})},\ \Eprint
  {http://arxiv.org/abs/1108.2138} {arXiv:1108.2138 [astro-ph.SR]} \BibitemShut
  {NoStop}%
\bibitem [{\citenamefont {Kalapotharakos}\ \emph
  {et~al.}(2012{\natexlab{b}})\citenamefont {Kalapotharakos}, \citenamefont
  {Harding}, \citenamefont {Kazanas},\ and\ \citenamefont
  {Contopoulos}}]{Kalapotharakos:2012dq}%
  \BibitemOpen
  \bibfield  {author} {\bibinfo {author} {\bibfnamefont {C.}~\bibnamefont
  {Kalapotharakos}}, \bibinfo {author} {\bibfnamefont {A.~K.}\ \bibnamefont
  {Harding}}, \bibinfo {author} {\bibfnamefont {D.}~\bibnamefont {Kazanas}}, \
  and\ \bibinfo {author} {\bibfnamefont {I.}~\bibnamefont {Contopoulos}},\
  }\href {\doibase 10.1088/2041-8205/754/1/L1} {\bibfield  {journal} {\bibinfo
  {journal} {Astrophys. J. Lett.}\ }\textbf {\bibinfo {volume} {754}},\
  \bibinfo {pages} {L1} (\bibinfo {year} {2012}{\natexlab{b}})},\ \Eprint
  {http://arxiv.org/abs/1205.5769} {arXiv:1205.5769 [astro-ph.HE]} \BibitemShut
  {NoStop}%
\bibitem [{\citenamefont {Bhattacharya}\ and\ \citenamefont
  {Pal}(2004)}]{Bhattacharya:2002qf}%
  \BibitemOpen
  \bibfield  {author} {\bibinfo {author} {\bibfnamefont {K.}~\bibnamefont
  {Bhattacharya}}\ and\ \bibinfo {author} {\bibfnamefont {P.~B.}\ \bibnamefont
  {Pal}},\ }\href {\doibase 10.1007/BF02705251} {\bibfield  {journal} {\bibinfo
   {journal} {Pramana}\ }\textbf {\bibinfo {volume} {62}},\ \bibinfo {pages}
  {1041} (\bibinfo {year} {2004})},\ \Eprint
  {http://arxiv.org/abs/hep-ph/0209053} {arXiv:hep-ph/0209053} \BibitemShut
  {NoStop}%
\end{thebibliography}
 
%

\end{document}